\documentclass[%
 reprint,
nofootinbib,
 amsmath,amssymb,
 aps,
]{revtex4-2}

\usepackage{amsmath}
\usepackage{graphicx}
\usepackage{xcolor}

\usepackage{hyperref}
\usepackage{mathtools}
\usepackage[caption=false]{subfig}

\newcommand{\beq}{\begin{equation}}
\newcommand{\eeq}{\end{equation}}
\newcommand{\bea}{\begin{eqnarray}}
\newcommand{\eea}{\end{eqnarray}}

\newcommand{\rev}[1]{{\color{black}#1}}

\usepackage[normalem]{ulem}

\begin{document}

\title{Extreme Axions Unveiled: a Novel Fluid Approach for Cosmological Modeling}

\author{Harrison\ Winch}\email{harrison.winch@mail.utoronto.ca}
\affiliation{Department of Astronomy \& Astrophysics, University of Toronto}
\affiliation{Dunlap Institute for Astronomy and Astrophysics, University of Toronto, 50 St. George Street, Toronto, ON M5S 3H4, Canada}

\author{Ren\'ee Hlo\v zek}
\affiliation{Department of Astronomy \& Astrophysics, University of Toronto}
\affiliation{Dunlap Institute for Astronomy and Astrophysics, University of Toronto, 50 St. George Street, Toronto, ON M5S 3H4, Canada}

\author{David J. E. Marsh}
\affiliation{Theoretical Particle Physics and Cosmology, King’s College London, Strand, London, WC2R 2LS, United Kingdom}

\author{Daniel Grin}
\affiliation{Haverford College, 370 Lancaster Ave, Haverford PA, 19041, USA}

\author{Keir K. Rogers}
\affiliation{Dunlap Institute for Astronomy and Astrophysics, University of Toronto, 50 St. George Street, Toronto, ON M5S 3H4, Canada}

\begin{abstract}
Axion-like particles (ALPs) are a well-motivated dark matter candidate that solve some of the problems in the clustering of large scale structure in cosmology. ALPs are often described by a simplified quadratic potential to specify the dynamics of the axion field, and are included in cosmological analysis codes using a modified fluid prescription. In this paper we consider the extreme axion: a version of the axion with a high initial field angle that produces an enhancement (rather than a suppression) of structure on small scales around the Jeans length, which can be probed by measurements of clustering such as the eBOSS DR14 Ly-$\alpha$ forest.  We present a novel method of modeling the extreme axion as a cosmological fluid, combining the Generalized Dark Matter model with the effective fluid approach presented in the \texttt{axionCAMB} software, as well as implementing a series of computational innovations to efficiently simulate the extreme axions. We find that for axion masses between $10^{-23} \mathrm{\ eV} \lesssim m_\mathrm{ax} \lesssim 10^{-22.5} \mathrm{\ eV}$, constraints on the axion fraction imposed by the eBOSS DR14 Ly-$\alpha$ forest can be significantly weakened by allowing them to be in the form of extreme axions with a starting angle between $\pi - 10^{-1} \lesssim \theta_i \lesssim \pi - 10^{-2}$. This work motivates and enables a more robust hydrodynamical analysis of extreme axions in order to compare them to high-resolution Ly-$\alpha$ forest data in the future.

\end{abstract}

\maketitle
\section{Introduction}

\label{sec:Intro} 

Axion-like particles (ALPs) are a broad class of dark matter (DM) particle candidates that possess both a strong theoretical justification and a variety of potentially observable signatures. While the traditional quantum chromodynamics (QCD) axion is a pseudo-Nambu-Goldstone boson arising from a broken Peccei-Quinn symmetry \cite{PecceiQuin:1977}, ALPs can arise from broken symmetries more generally, and are produced naturally from a variety of string theories as a result of compactified higher dimensions, making them a well-motivated DM particle candidate \cite{Dine:1982ah,Preskill:1982cy,Abbott:1982af,Svrcek:2006yi, Duffy:2009ig,Arvanitaki:2009fg,  Marsh:2015xka,Adams:2022pbo}. Throughout this work, we will use axion and ALP interchangeably to refer to this broad class of low-mass pseudo-Nambu-Goldstone boson DM candidates.

The extremely flat field potential of ALPs gives them a very low particle mass $m_\mathrm{ax}$, potentially on the order of $10^{-22}$ eV. This extremely low mass results in a de-Broglie wavelength that ``smooths'' cosmic power on small ($\sim$ kpc - Mpc) scales, which is why these models are sometimes called ``fuzzy dark matter'' \cite{Hu:2000ke, Dentler:2021zij}. The scale of this power suppression is directly related to the ALP mass, with lower masses suppressing structure on larger scales \cite{Lague:2020htq}. This has allowed us to put strong lower bounds on the axion mass (or upper bounds on the axion fraction), using a variety of observables such as galaxy clustering, the Lyman-$\alpha$ forest, and the Cosmic Microwave Background (CMB) \cite{Amendola:2005ad,Hlozek:2016lzm,Hlozek:2017zzf, Rogers:2020ltq, Dentler:2021zij}. For a more detailed review of axions and their role in cosmology, see, e.g., Refs. \onlinecite{Marsh:2015xka} and \onlinecite{Grin:2019mub}. 

However, most analyses of dark matter structure formation ignore the periodic nature of the Nambu-Goldstone field, which creates a cosine field potential for the axion. While the mass of the ALP $m_\mathrm{ax}$ characterizes the field potential curvature at a stable minimum, the scale of the periodicity of the field potential is related to the energy scales of the broken symmetry giving rise to the Nambu-Goldstone field, parameterized by the axion decay constant $f_\mathrm{ax}$ \cite{Marsh:2015xka}. We can write the field potential $V(\phi)$ for the axion field $\phi$ as
\begin{equation}
    V(\phi) = m_\mathrm{ax}^2 f_\mathrm{ax}^2 \big[1 - \cos(\phi / f_\mathrm{ax}) \big],
    \label{eq:cosine_pot}
\end{equation}
with a stable minimum at $\phi = 0$, and periodicity on the order of $\phi \sim 2 \pi f_\mathrm{ax}$. If the axion field exhibits only small oscillations around this minimum, we can approximate the potential near the minimum as quadratic, of the form
\begin{equation}
    V(\phi) \approx \frac{m_\mathrm{ax}^2}{2} \phi^2.
    \label{eq:quad_pot}
\end{equation}
All dependence on the symmetry-breaking scale $f_\mathrm{ax}$ cancels out, and we get a harmonic potential with curvature depending on the axion mass $m_\mathrm{ax}$. 

Modeling a scalar field in a harmonic potential is a good approximation of the ALP behaviour both at late times or if you assume a low starting field value relative to $f_\mathrm{ax}$. As a result, most past work in axion cosmology has assumed a low starting angle and thus a purely quadratic potential, allowing the ALP field to be approximated as a generalized dark matter fluid, and efficient predictions can be made of cosmological observables (i.e., the suppression in small-scale power mentioned earlier) \cite{Hu:1998kj, PhysRevD.91.103512, Lague:2020htq}.

However, despite the computational simplicity of the quadratic potential, the full cosine nature of the potential becomes significant if you start the axion field near the ``top'' of the potential (so $\theta_i \equiv \phi_i / f_\mathrm{ax} \rightarrow \pi$, where $\theta_i$ is the initial value of the axion field angle, $\theta$). In this work, we will refer to models where the axion field starts near the top of the cosine potential as ``extreme'', following the convention in existing literature labeling these models as ``extreme axion dark matter'' or ``extreme wave dark matter'' \cite{Cedeno:2017sou, Zhang:2017dpp}. \rev{These models are sometimes also refered to as ``large-misalignment angle" axions \citep[eg.][]{Arvanitaki:2019rax}.} In contrast to these ``extreme'' models, we will refer to axions with low starting angles well within the quadratic regime as ``vanilla'' axions.

Past work modeling perturbations to the axion field have shown that starting near the top of the potential results in significant enhancements to the Matter Power Spectrum (MPS) around the same scales that are ordinarily suppressed in the vanilla axion case \cite{Cedeno:2017sou, Leong:2018opi, LinaresCedeno:2020dte}. This enhancement arises in the scalar field as it evolves over a region of field potential with negative curvature\footnote{For an intuitive understanding of why negative potential curvature leads to an enhancement of power, consider the following argument: if the curvature of the field potential is related to $m_\mathrm{ax}^2$ in the case of a free particle, negative curvature is a negative $m_\mathrm{ax}^2$ term, and thus can be thought of as ``imaginary mass'', as the square root of a negative number. Since the mass governs the frequency of the field oscillations, this gives rise to an ``imaginary frequency''. Oscillations with an imaginary frequency are just hyperbolic sine and cosine functions, so the field perturbations exhibit exponential growth instead of harmonic oscillation during this regime of negative potential curvature.}. These enhancements have been shown to weaken, or even reverse, the suppression of power due to the low axion mass, resulting in a weakening of existing observational constraints on these axion models. The scale of these enhancements is related to how close the axion field angle, $\theta_i,$ starts to $\pi$, as this results in the field remaining ``balanced'' on the top of the potential for longer, and spending more time in this region of negative potential curvature. Thus, developing ways of efficiently modeling these extreme axions can allow us to reevaluate the robustness of past axion constraints, and explore new and interesting models of DM that are still consistent with the data.


Past work has suggested this ``balanced'' starting value requires some degree of fine-tuning of the initial conditions, but models have been proposed that could explain this fine-tuning, with an inverted potential at an earlier phase driving the axion field to start near the potential maximum at $\phi \sim 2 \pi f_\mathrm{ax}$ \cite{Co:2018mho, Arvanitaki:2019rax}. Being able to efficiently model the extreme axions for a range of parameters, and comparing the results to cosmological observables, would allow us to test the required degree of fine-tuning of the initial field angle, and thus put constraints on the range of possible axion models.

Some work has been done to model the evolution of the axion field with these extreme starting angles \cite{Cedeno:2017sou, Zhang:2017flu, Leong:2018opi, Zhang:2017dpp}. However, the rapidly-oscillating nature of these axion fields (both at the background and perturbation level) necessitates extremely high temporal resolution for the computations, requiring long computation times for a brute force solution \cite{Zhang:2017flu,Zhang:2017dpp}. This makes running repeated estimates of the axion evolution, of the sort required for a Markov Chain Monte Carlo (MCMC) or other likelihood sampler method, prohibitively expensive. 

In this work, we present a novel method of efficiently and accurately modeling the behaviour of these extreme axions as a cosmological fluid. We follow the structure of the vanilla axion modeling code \texttt{axionCAMB}, explained in more detail in Ref. \onlinecite{PhysRevD.91.103512}. We implement a number of innovations and improvements to \texttt{axionCAMB} compute predictions for cosmological observables such as the linear MPS. These innovations, described in more detail in the Section~\ref{sec:methods}, include a restructuring of the initial conditions, and a novel effective sound speed of the extreme axion fluid. All of these innovations reduce the run time to model extreme axions down to $\sim 7$ seconds. This opens up new opportunities to put observational constraints on extreme axion models with higher-dimensional MCMC algorithms that require tens of thousands of calls to the axion evolution code. \rev{Our code is shared on a Github repository: \href{https://github.com/HarrisonWinch96/AxionCambRenIso_Extreme.git}{https://bit.ly/axionCAMBExtreme}.}

In addition to explaining our various novel innovations to model extreme axions as a cosmological fluid, we also present predictions for some cosmological observables, in order to assess the potential of these models to be observed. We compare our predictions for the linear matter power spectrum to estimates of the linear MPS from the the Extended Baryonic Oscillation Spectroscopic Survey (eBOSS) DR14 Ly-$\alpha$ forest \cite{eboss}. We find that moderately extreme axions can alleviate tensions between vanilla axions and the Ly-$\alpha$ estimates for a range of masses and axion fractions, and that the improvements in the fit are significant enough to warrant the addition of the extra parameter to our vanilla axion model. For example, we show that for axion masses between $10^{-23} \text{ eV} \lesssim m_\mathrm{ax} \lesssim 10^{-22.5} \text{ eV}$, constraints on the axion fraction imposed by the eBOSS DR14 Ly$\alpha$ forest can be significantly weakened by considering extreme axions with a starting angle between $\pi - 10^{-1} \lesssim \theta_i \lesssim \pi - 10^{-2}$. This motivates future hydrodynamical simulations of the Ly-$\alpha$ forest in extreme axion cosmologies, in order to compute more robust comparisons to these models, similar to the analysis done in Ref. \onlinecite{Rogers:2020ltq}.

\section{Methods}
\label{sec:methods}
In order to model the behavior of extreme axions, we modified \texttt{axionCAMB} to include an arbitrary field potential shape (in our case, a cosine of the form given in Equation \ref{eq:cosine_pot}), and reconfigured the code to sample the extreme starting angles necessary to probe these potentials. We also modified the effective sound speed of the axions after the onset of oscillations to reflect the growth in structure resulting from the tachyonic field dynamics. Lastly, we implemented a computationally-efficient `lookup table' of the axion background fluid evolution in order to speed up the computation of the perturbation equations of motion. The details of implementing extreme axions into \texttt{axionCAMB} are presented below \footnote{\texttt{axionCAMB} is in turn based on the cosmological Boltzmann code, \texttt{CAMB} \cite{Lewis:2002ah}.}. 

\subsection{Review of \texttt{axionCAMB}}

The numerical treatment of axions in \texttt{axionCAMB} is described in detail in Ref. \onlinecite{PhysRevD.91.103512}, but we review the dynamics of axions here in a potential-agnostic way in order to set up our discussion of modeling extreme axions. In theory, the best way of modeling the dynamics of axion dark matter is to model the behaviour of the field throughout all of cosmic history, and derive all cosmological parameters from those primary variables. However, since this field evolution includes periods of extremely rapid oscillations at late times, simulating this is computationally prohibitive and numerically unstable. Instead, the axion field is modeled directly at early times, but the code switches to a simplified fluid approximation at late times \cite{PhysRevD.91.103512}. This piece-wise background evolution could then be called when solving the equations of motion for the fluid perturbations (axion density perturbation $\delta_\mathrm{ax}$ and axion heat flux $u$), allowing for efficient and stable computation of the final axion power spectrum. This method is discussed here, building on the discussion in Refs. \onlinecite{Hu:1998kj} and \onlinecite{PhysRevD.91.103512}. 

The axion field, $\phi$, can be broken up into a background ($\phi_0$) and perturbation ($\phi_1$) component, so 
\begin{equation}
    \phi(\tau, k) = \phi_0(\tau) + \phi_1(\tau, k),
\end{equation}
where $\tau$ is conformal time, and $k$ is spatial wavenumber. Conformal time can be computed from scale factor $a$ and Hubble parameter $H(a)$ using the formula
\begin{equation}
    \tau = \int \frac{da}{a^2 H(a)}.
\end{equation}

The equation of motion of a background axion field $\phi_{0}$ can be computed from the field Lagrangian, and has the form
\begin{equation}
    \ddot{\phi}_{0} + 2 \mathcal{H} \dot{\phi}_{0} + a^2 V'(\phi_{0}) = 0,
    \label{eq:background_EoM}
\end{equation}
where $\phi_{0}$ is the background axion field, $\mathcal{H}$ is the conformal Hubble parameter, $a$ is the scale factor, $V'(\phi_0)$ is the field derivative of the potential $V(\phi_0)$ and the dots represent derivatives with respect to conformal time $\tau$. In theory, the field potential $V(\phi)$ could be any functional form, but the canonical axion-like particle model has a cosine potential of the form
\begin{equation}
V(\phi) = m_\mathrm{ax}^2 f_\mathrm{ax}^2 [1 - \cos(\phi/f_\mathrm{ax})],
\end{equation}
where $m_\mathrm{ax}$ is the axion mass, and $f_\mathrm{ax}$ is the symmetry-breaking scale, also referred to as the axion decay constant 
\footnote{Most treatments of the axion, including vanilla \texttt{axionCAMB}, assume that the field is in the minimum of this potential. This allows them to approximate the cosine potential as a quadratic with curvature proportional to the axion mass as $ V(\phi) \approx m_\mathrm{ax}^2 \phi^2/2. $}.

The background density and pressure of the axion field can also be derived from the Lagrangian, and can be expressed as
\begin{align}
    \rho_\mathrm{ax} &= \frac{a^{-2}}{2}\dot{\phi}_{0}^2 + V(\phi_{0}) \\
    P_\mathrm{ax} &= \frac{a^{-2}}{2}\dot{\phi}_{0}^2 - V(\phi_{0}),
\end{align}

The equation of motion given by Eq. \ref{eq:background_EoM} results in an axion background field that remains fixed at early times ($3\mathcal{H}/a \gtrsim m_\mathrm{ax}$), behaving like a cosmological constant with fixed density, but begins to  oscillate rapidly at or soon after $3\mathcal{H}/a \sim m_\mathrm{ax}$, as the Hubble friction term becomes subdominant. Eventually, the field settles in a local minimum, and evolves like a free-particle, well approximated by the quadratic potential. At this point, the density decays like $a^{-3}$, analogous to Cold Dark Matter (CDM).


\texttt{axionCAMB} runs through this early pre-oscillatory phase several times in order to determine the proper initial value of the axion field required to produce the desired final axion density, and the time at which it can safely switch back to the free particle CDM solution at late times. It then dynamically evolves these initial conditions \citep[integrating the equation of motion for a field using a Runge-Kutta integrator,][]{Runge1895UeberDN} until the field starts to oscillate, at which point it switches over to the known free-particle solution for DM evolution \cite{PhysRevD.91.103512}. 

From these primary background fluid variables, secondary background fluid variables are computed, which can be used directly in the density perturbation equations of motion. These secondary background fluid variables are the equation of state parameter $w_\mathrm{ax}$, and the adiabatic sound speed squared $c_\mathrm{ad}^2$, given by the formulae
\begin{align}
    \label{eqn:wax}
    w_\mathrm{ax} &= \frac{P_\mathrm{ax}}{\rho_\mathrm{ax}} \\
    \label{eqn:cad2}
    c_\mathrm{ad}^2 &= \frac{\dot{P}_\mathrm{ax}}{\dot{\rho}_\mathrm{ax}} = w_\mathrm{ax} - \frac{\dot{w}_\mathrm{ax}}{3 H (1 + w_\mathrm{ax})}.
\end{align}

These secondary background variables ($w_\mathrm{ax}$ and $c_\mathrm{ad}^2$) are then used in \texttt{axionCAMB} to compute the equations of motion for the axion density perturbations in the comoving synchronous gauge, which are exact at early times. For an axion density perturbation $\delta_\mathrm{ax}$ and heat flux $u_\mathrm{ax}$ of wavenumber $k$, the equations of motion can be expressed as the pair of first-order differential equations,
\begin{align}
    \dot{\delta}_\mathrm{ax} &= - {k} u_\mathrm{ax} - (1 + w_\mathrm{ax}) \dot{\beta}/2 - 3H(1 - w_\mathrm{ax})\delta_\mathrm{ax} \nonumber \\
    & \,\,\,- 9 H^2(1 - c_\mathrm{ad}^2)u_\mathrm{ax}/{k} \\
    \dot{u}_\mathrm{ax} &= 2 H u_\mathrm{ax} + {k} \delta_\mathrm{ax} + 2 H (w_\mathrm{ax} - c_\mathrm{ad}^2) u_\mathrm{ax}.
\end{align}
Here, $w_\mathrm{ax}$ and $c_\mathrm{ad}^2$ are the secondary axion background variables from above, and $\beta$ is the trace of the scalar metric perturbation in synchronous gauge (providing a gravitational driving term for these perturbation oscillations).

After the onset of oscillations, the background fluid variables of $w_\mathrm{ax}$ and $c_\mathrm{ad}^2$ can both be approximated as zero in the case of a quadratic potential, analytically averaging over the rapid oscillations of the axion field \cite{PhysRevD.91.103512}. In addition, after the onset of oscillations, the axion sound speed $c_\mathrm{ax}^2$ (not to be confused with the \textit{adiabatic} sound speed $c_\mathrm{ad}^2$) can no longer be set to unity \cite{Khlopov:1985jw} through the choice of comoving gauge, and instead becomes approximated by the k-dependent expression \cite{Hwang:2009js}: 
\begin{equation}
\label{eq:cax_approx}
    c_\mathrm{ax}^2 \equiv \frac{\delta P}{\delta \rho} \approx \frac{k^2/(4 m_\mathrm{ax}^2 a^2)}{1 + k^2/(4 m_\mathrm{ax}^2 a^2)}.
\end{equation}
This results in a new set of equations of motion for the axion perturbations after the onset of oscillations:
\begin{align}
    \label{eqn:pert_EoM}
    \dot{\delta}_\mathrm{ax} &= -k u_\mathrm{ax} - \frac{\dot{\beta}}{2} - 3 H^2 c_\mathrm{ax}^2 \delta_\mathrm{ax} - 9 H^2 c_\mathrm{ax}^2 u_\mathrm{ax}/k \\
    \dot{u}_\mathrm{ax} &= -Hu_\mathrm{ax} + c_\mathrm{ax}^2 k \delta_\mathrm{ax} + 3 c_\mathrm{ax}^2 H u_\mathrm{ax}
\end{align}
The perturbation equations of motion in these two regimes can be used to calculate the evolution of the axion perturbations, and make predictions for cosmological observables such as the MPS or CMB.

\subsection{Finely tuned initial conditions}
\label{sec:init}
In the original formulation of \texttt{axionCAMB}, the initial field angle was found by testing a range of starting field values $\phi_{i}$ in a fixed potential, evolving them all forward in time to find the final axion density, and then interpolating to find the initial field value that best reproduces the desired final density via this ``shooting method''. However, in order to explore the tachyonic enhancements arising from extreme starting angles, we need to set the starting angle ($\theta_\mathrm{ax} = \phi_\mathrm{ax}/f_\mathrm{ax}$) extremely close to $\pi$, which is impossible to do manually in the original formulation of \texttt{axionCAMB}. 

\begin{figure}
	\centering
    \includegraphics[width=0.5\textwidth]{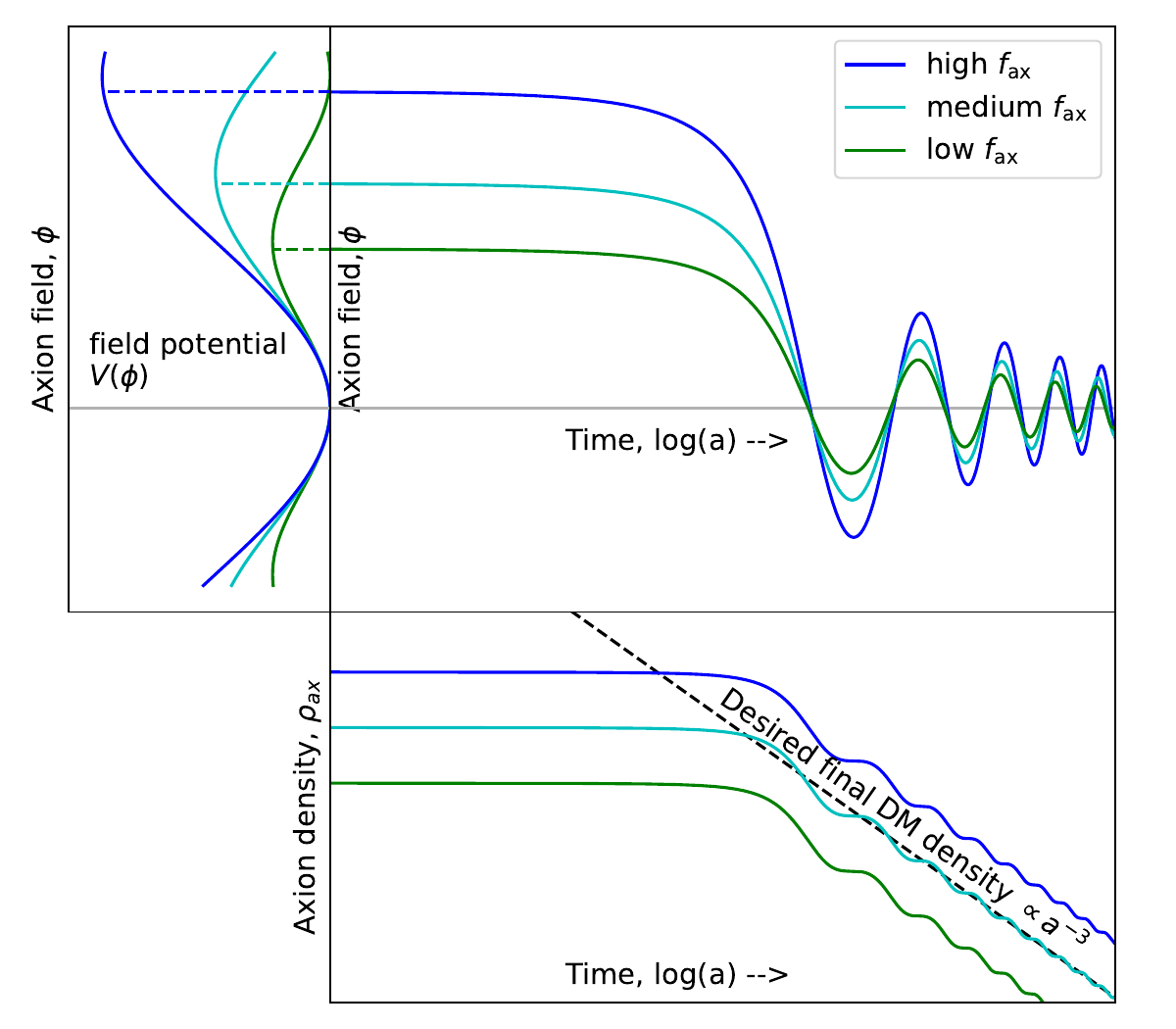}
   \caption{This diagram illustrates our novel shooting method for determining the axion initial conditions, as explained in Section~\ref{sec:init}. Three possible axion potential scales are shown here in blue, cyan, and green. The initial value of the axion field is determined by the axion starting angle (in radians, set here to be 3.0), and the scale of the axion potential. We then evolve the axion field forward in time using equation \ref{eq:background_EoM}, as it starts to oscillate at late times, as shown in the top panel. Once the axion density has started to evolve like CDM, we can compare the final densities of all of these test cases to the desired final axion density (shown in a black dashed line in the bottom-right plot), and we use a cubic spline interpolation to determine the correct potential scale to reproduce the desired final density.}  
    \label{fig:shooting}
\end{figure}

In order to specify both the initial axion field angle $\theta_{i}$ \textit{and} the final axion density $\Omega_{\mathrm{ax}0}h^2$, we had to restructure the initial shooting regime. In our new regime, the free parameter in the shooting regime is the scale of the cosine field potential, $f_\mathrm{ax}$. We test a range of field potential scales for a fixed axion mass $m_\mathrm{ax}$, starting all of the axion fields at the same \textit{angle} ($\phi_{i} = f_\mathrm{ax} \theta_{i}$) within the cosine potential, and then evolve them all forward in time using Equation \ref{eq:background_EoM} to find the final axion density. This process is illustrated in Figure~\ref{fig:shooting}. We are then able to interpolate from these final densities in order to find the field potential scale that correctly reproduces the desired final axion density given a certain starting angle.

Restructuring the initial shooting methods to specify the field starting angle allows us to probe the effects of extreme starting angles in new ways. We can specify starting angles arbitrarily close to $\pi$, in order to see the effects of these extremely finely tuned angles on other observables. In addition, when performing MCMC analysis, having the starting angle as a free parameter allows us to impose arbitrary priors on this starting angle. We can use these priors to test the dependence of any constraints on the level of fine-tuning of the axion starting angle.

\subsection{Modeling the early-oscillatory effective axion sound speed}
\label{sec:sspeed}

While the pre-oscillatory behaviour of the extreme axion perturbations can be precisely modeled by the equation of state parameter $w_\mathrm{ax}$ and the adiabatic sound speed $c_\mathrm{ad}^2$, this is no longer the case after the onset of oscillations. This is because the adiabatic sound speed, defined by Equation \ref{eqn:cad2}, becomes undefined when $\dot{\rho}_\mathrm{ax} = 0$, or when $w_\mathrm{ax} = -1$, and these poles are difficult to integrate around. Therefore, after the onset of oscillations, we must switch over to an effective-fluid formalism, where we time-average over the oscillations. This gives us $w_\mathrm{ax} = 0$ and $c_\mathrm{ad}^2 = 0$, and the behaviour of the perturbation equations of motion are instead governed by the effective axion sound speed $c_\mathrm{ax}^2$, as seen in Eq. \ref{eqn:pert_EoM}. This sound speed is an approximation, which, in the case of harmonic oscillations of a vanilla axion, can be approximated analytically to be, the expression given in Eq. \ref{eq:cax_approx}.

However, in the case of an extreme axion, particularly during the early anharmonic phase of oscillations, the assumptions of regular harmonic oscillations do not hold. The nature of anharmonic oscillations resists easy analytic approximation of the effective fluid sound speed, so instead we developed methods of approximating this sound speed numerically. Since the effective fluid sound speed describes the frequency of oscillations, exponential growth of the axion perturbations can be modeled by a negative value of $c_\mathrm{ax}^2$, much like the negative mass squared term in the field equations. In this case, a negative $c_\mathrm{ax}^2$ term does not actually mean space-like sound speeds, but instead is merely an effective fluid model of the axion field instabilities. The impact of this negative sound speed on the growth of the fluid perturbations is roughly proportional to the integral of the sound speed over conformal time.

\begin{figure}
	\centering
    \includegraphics[trim={0.2cm 1.2cm 1cm 2.5cm},clip,width=0.5\textwidth]{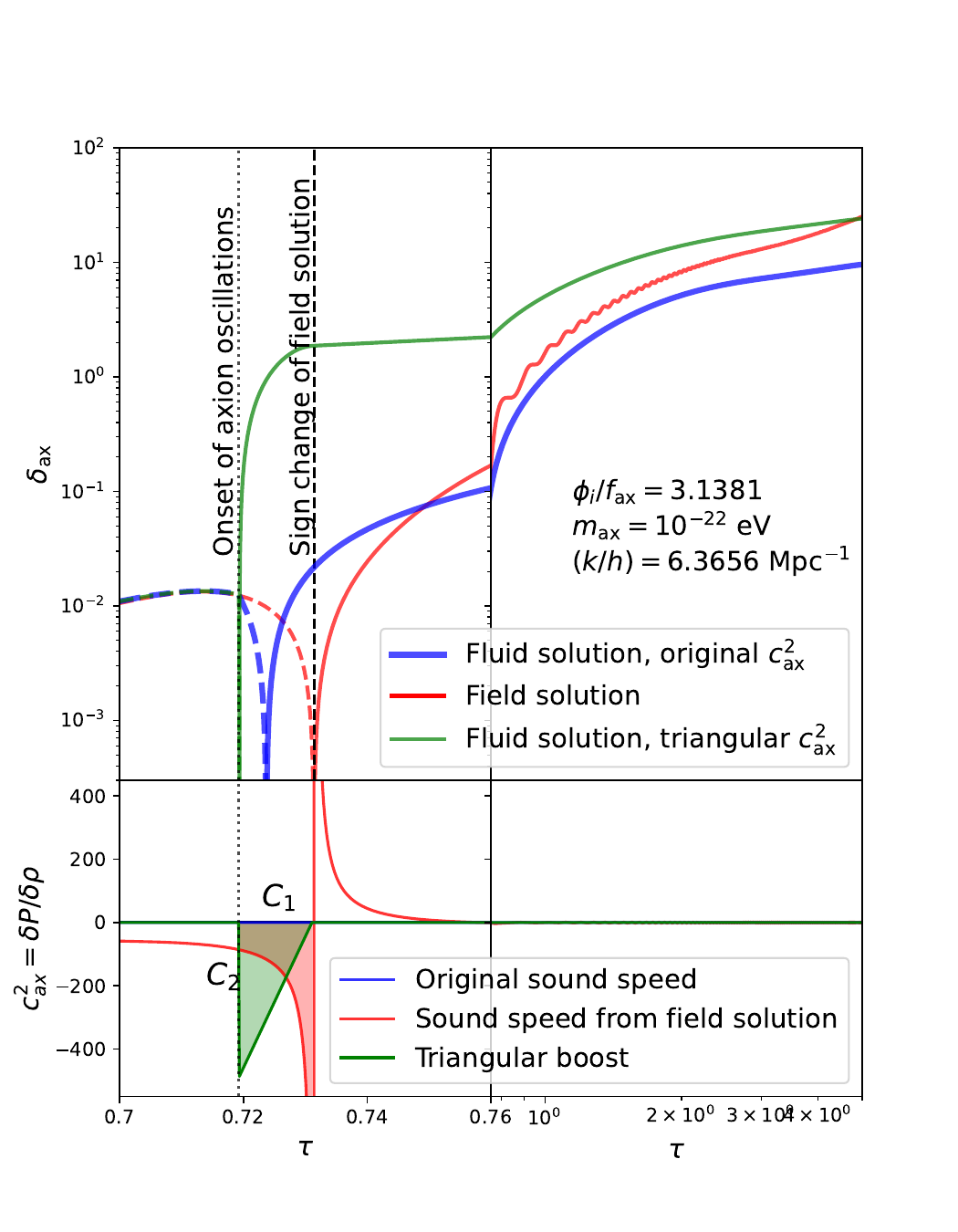}
   \caption{The evolution of axion perturbations during and soon after the onset of oscillations for different treatments of the axion sound speed. The horizontal axis is divided between a linear scale on the left and a log scale on the right, to capture both the early and late time behaviour. The blue line represents the \rev{generalized dark matter (GDM)} fluid equations with the default vanilla axion sound speed \cite{Hu:1998kj} The red line in the upper plot uses the Klein-Gordon field equations with the metric term sourced from the vanilla fluid solution. The red line in the lower plot shows the axion sound speed computed from this field solution ($c_\mathrm{ax}^2 = \delta P / \delta \rho$) in the synchronous gauge. This sound speed is used to fit the approximate height and width of a triangular boost to this sound speed - this boost is shown in green on the lower plot. In the upper plot, the green curve shows the solution to the fluid GDM equations with this triangular boost in the sound speed, reproducing the expected power at late times.}  
    \label{fig:cax_explanatory}
\end{figure}

To get a sense of the effects of the anharmonic potential on the axion fluid sound speed, we first solve the axion field perturbation equations of motion, 
\begin{equation}
    \ddot{\phi}_1 + 2 \mathcal{H}\dot{\phi}_1 + \big[k^2 + a^2V''(\phi_0)\big]\phi_1 = -\frac{1}{2}\dot{\phi_0}\dot{\beta},
\end{equation}
where $\phi_1 = \phi - \phi_0$ is the axion field perturbation. For this integration, we use the metric term $\dot{\beta}$ from the \texttt{axionCAMB} solution with the vanilla axion sound speed as the driving source term. This approach is accurate in the adiabatic mode before equality when radiation dominates the gravitational potential. We can then use this $\{ \phi_0, \dot{\phi}_0, \phi_1, \dot{\phi}_1 \}$ solution to compute the fluid sound speed in synchronous gauge,
\begin{equation}
    c_\mathrm{ax}^2 = \frac{\delta P_\mathrm{ax}}{\delta \rho_\mathrm{ax}} = \frac{a^{-2}\dot{\phi}_0 \dot{\phi}_1 - V'(\phi_0) \phi_1}{a^{-2}\dot{\phi}_0 \dot{\phi}_1 + V'(\phi_0) \phi_1}.
\end{equation}
This approximation of the fluid sound speed is shown in red in the lower subplot of Figure~\ref{fig:cax_explanatory}.

In order to approximate the boost in the axion sound speed shown in the field equations without changing the late-time evolution of the perturbations, we modified the vanilla axion fluid sound speed to include a large negative spike just after the onset of oscillations. This negative triangular spike is shown in green in the lower subplot of Figure~\ref{fig:cax_explanatory}. The width and height of this spike were fit to match the approximate sound speed computed from the field perturbation solution. The width ($C_1$) was fit to the delay in scale factor $a$ between the onset of axion oscillations and the asymptotic sign change in the field solution sound speed. This numerical width was then approximated as a power law function of the scale factor $k$ of the perturbation, depends linearly on the scale factor at the onset of oscillations, which in turn depends on the axion mass, fraction, and starting angle,
\begin{equation}
    C_1(k, a_\mathrm{osc})_{\{A_1, M_1, B_1\}} \approx \frac{a_\mathrm{osc}}{A_1}\bigg(\frac{k}{B_1}\bigg)^{M_1},
\end{equation}
\rev{where $A_1$, $M_1$, and $B_1$ are all fit parameters, chosen to match the dependance of the scale factor delay in the field solutions over a range of values for $k$ and $a_\mathrm{osc}$.}

The height of this triangle ($C_2$)was chosen such that the total area of the triangle was equal to the area enclosed by the sound speed over the same conformal time period. As mentioned earlier, since the axion equations motion are differential equations in conformal time $\tau$, the integral of the $c_\mathrm{ax}^2$ over $\tau$ approximates the impact that this sound speed spike has on the evolution of the perturbations. Once this height was calculated for a certain field evolution, it is effectively fit by an power law of the sale factor $k$ and the logarithm of the initial axion field angle separation from $\pi$ (since more extreme starting angles should result in exponentially larger boosts to the perturbation growth)\footnote{A logarithmic dependence for the background field can be derived analytically for the anharmonic corrections to the relic density~\cite{Lyth:1991ub}.},
\begin{equation}
    C_2(k, \theta_{\mathrm{ax},i})_{\{Q_2, M_2, B_2\}} \approx \bigg[\log\bigg(\frac{\pi - \theta_{\mathrm{ax},i}}{Q_2}\bigg)\bigg]\bigg(\frac{k}{B_2}\bigg)^{M_2},
\end{equation}
\rev{where $Q_2$, $M_2$, and $B_2$ are all fit parameters, chosen to match the area of the effective sound speed calculated from the field solutions over a range of $k$ and $\theta_{\mathrm{ax},i}$ values.}

\begin{figure}
	\centering
    \includegraphics[width=0.5\textwidth]{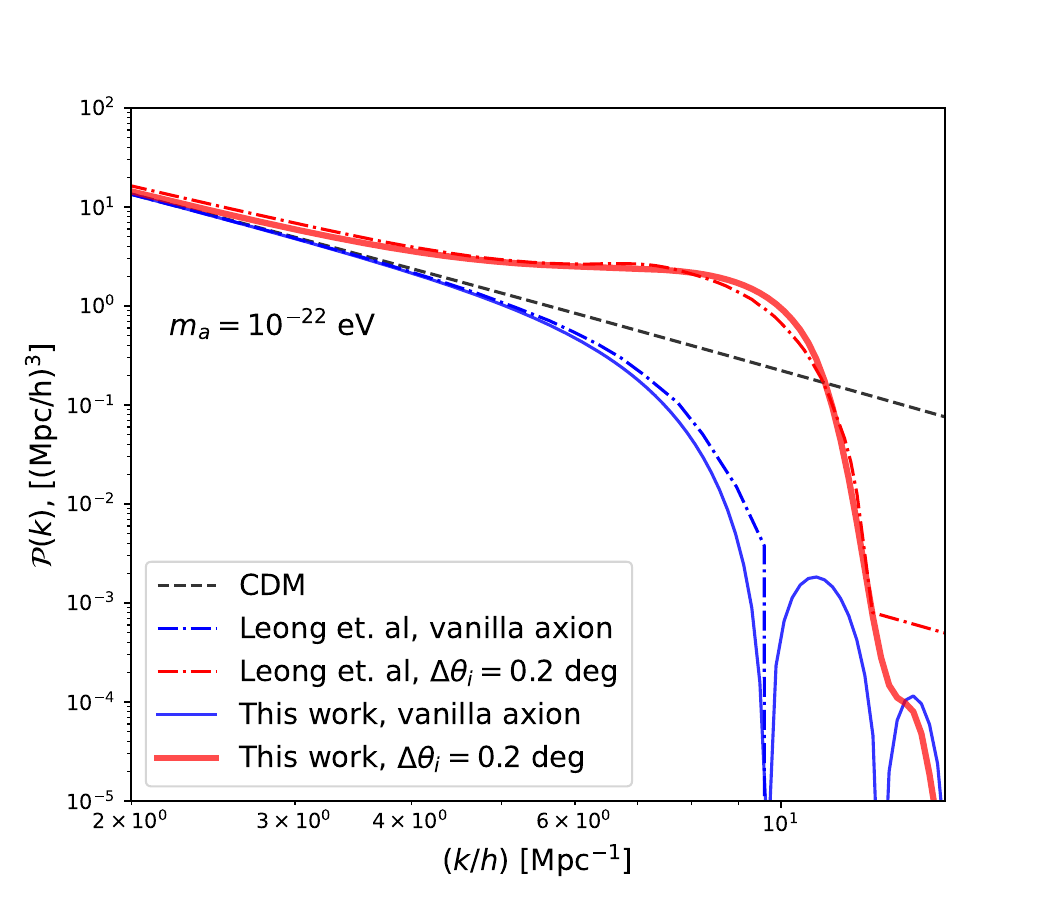}
   \caption{This figure compares the predicted matter power spectra for our technique of fitting a triangular boost to the axion sound speed, to that predicted in Ref. \onlinecite{Leong:2018opi}, which used the full field perturbation equation solution to compute the matter power spectrum for an extreme axion.}  
    \label{fig:leong_comparison}
\end{figure}

With these two functions for $C_1$ and $C_2$, we can compute a sound speed boost for any combination of axion parameters, approximating the effect of the fluid sound speed for a rapidly oscillating extreme axion. This modifies the vanilla axion sound speed with the following triangular boost,
\begin{equation}
    c_\mathrm{ax}^2 = \tilde{c}_\mathrm{ax}^2 - \bigg[ C_2* \frac{(a_\mathrm{osc} + C_1) - a}{C_1} ; a \in (a_\mathrm{osc}, a_\mathrm{osc} + C_1)\bigg]
\end{equation}
The power spectrum results for this method can be compared to the literature, where other groups have used the exact field perturbation equations of motion to compute the matter power spectrum for extreme axions, such as Ref. \onlinecite{Leong:2018opi}. In Figure~\ref{fig:leong_comparison} we can see the comparison in the matter power spectrum for both a vanilla axion and an extreme axion with a starting angle deviating from $\pi$ by $0.2$ degrees, and we find that they are in remarkably close agreement with Ref. \onlinecite{Leong:2018opi}. However, this close agreement seems to hold best at $z=0$, when these power spectra are computed, while the higher redshift comparison may be more nuanced. Figure~\ref{fig:cax_explanatory} suggests that while the exact field solution and the new approximate fluid solution agree at very late times, their evolution at early times are not fully equivalent, so more work may need to be done on this approximation in order to perform comparisons to high-redshift observables.

\subsection{Using lookup tables for efficient modeling of field}

Extending the full field evolution so much later past the onset of oscillations requires far more computational resources than ending the field evolution as soon as oscillations begin. Greater numerical resolution, in both time and possible field potential scale, is also required to integrate these rapidly oscillating variables. With these increases to computational time, the new version of \texttt{axionCAMB} takes around seventy seconds to complete. While this may be feasible when computing a single power spectrum result, this is computationally intensive with which to run an MCMC analysis, which may require tens to hundreds of thousands of separate calls to \texttt{axionCAMB}.

Fortunately, there is an opportunity here to streamline the process through the use of a pre-computed lookup table of smoothed axion background variables. The background evolution of the axion depends only on the axion mass $m_\mathrm{ax}$, axion density $\Omega_\mathrm{ax}h^2$, and axion starting angle $\theta_i$. In turn, the only output from the axion background module that is used by the rest of the code are the arrays of $w_\mathrm{ax}$, $c_\mathrm{ad}^2$, and $\rho_\mathrm{ax}$. 

We generated a lookup table of \texttt{axionCAMB} results $-24 < \log(m_\mathrm{ax})< -22$, $0.0012025 < \Omega_\mathrm{ax}h^2 < 0.12025$, and $\pi - 1 < \theta_i < \pi - 10^{-4}$, saving the arrays of the three background variables presented above, in addition to $\log(a)$ as a time variable. We were then able to write a new version of the axion background module that, instead of computing the axion background evolution from scratch, computes it instead from this lookup table. When this new version of \texttt{axionCAMB} is called, the new values of $m_\mathrm{ax}$, $\Omega_\mathrm{ax}h^2$, and $\theta_{i}$ are used to determine the eight reference combinations closest to the desired values. The proximity of the new values to these eight reference values is used to calculate a weighted average background evolution of $w_\mathrm{ax}$, $c_\mathrm{ad}^2$, $\rho_\mathrm{ax}$, and $\log(a)$. This lookup table method was tested extensively against the full computation, showing consistent results, and the required run time was reduced from $\sim 70$ seconds down to $\sim 7$ seconds.


\subsection{Summary of changes to \texttt{axionCAMB}}

In order to model axions with extreme starting angles in a cosine field potential using the computationally efficient field formalism used in \texttt{axionCAMB}, we have introduced a number of modifications to \texttt{axionCAMB} which are explained above, but summarized here. 
\begin{itemize}
    \item We replaced the quadratic approximation of the field potential with an arbitrary potential function, currently set to the canonical cosine potential.
    \item We restructured the initial conditions to specify the starting angle relative to this cosine, as well as the desired final axion density, and test a variety of potential scales $f_\mathrm{ax}$ to determine the correct one using a shooting method.
    \item We modified the effective axion fluid sound speed to reproduce the growth in structure seen in the exact field perturbation equations of motion.
    \item We pre-computed a lookup table for the axion background evolution which significantly reduced the run time.
\end{itemize}  
The result is an accurate modeling of extreme axion background and perturbation evolution for an arbitrary axion mass, density, and starting angle that only takes $\sim 7$ seconds to run. This powerful tool can shed new light on the behaviour and detectability of these extreme axion models, as discussed below.

\subsection{Data: Ly-$\alpha$ forest estimates of the MPS}
\label{sec:lya}

We introduce the data products we use to compare the observable effect of our extreme axion models, and to constrain these enhancements on small scales of the matter power spectrum. We consider the Ly-$\alpha$ forest data from the extended Baryonic Oscillation Spectroscopic Survey experiment \cite[eBOSS][]{eboss}. The Ly-$\alpha$ forest uses the absorption of light from high-redshift quasars by foreground neutral hydrogen. The neutral hydrogen absorbs light with the Lyman-$\alpha$ transition at rest wavelengths of $\lambda_{\text{Ly-}\alpha} = 121.6$ nm, but depending on the redshift at which this absorption occurs, the absorption feature will be detected at different places in the quasar spectrum. This allows one to sample the neutral hydrogen density along the entire line of sight to the quasar. This can result in extremely high-resolution estimates of the matter power spectrum, if we use a high number of quasars and high-resolution spectroscopy to analyze the quasar spectra. 

 We use the estimates of the linear MPS from the eBOSS DR14 Ly-$\alpha$ forest data, as this allows for direct comparison to our MPS predictions. The data are derived from  210,005 quasars with $z_q > 2.10$ that are used to measure the signal of Ly-$\alpha$ absorption \cite{eboss}. 
 The flux power spectrum is then used to compute estimates of the linear MPS at $z=0$, which we can compare to our MPS predictions\footnote{It should be noted that these estimates assume $\Lambda$CDM in their reconstruction, and that degeneracies might exist between the axion and astrophysical parameters used in the reconstruction. A more rigorous comparison between extreme axions and measurements of the Ly-$\alpha$ forest is left for future work, and discussed in Sec. \ref{sec:discussion}}.  These estimates of the linear MPS are shown in Figures \ref{fig:mps_theta}, \ref{fig:mps_mass}, and \ref{fig:mps_frac}, and a more robust comparison of many models is shown in Figures \ref{fig:chisgrid_23} and \ref{fig:chisgrid_225}. A more complete discussion of these results is presented in Section~\ref{sec:pheno}.


\section{Phenomenology}
\label{sec:pheno}
We discuss the observed changes to the axion background variables and cosmological observables as a result of the cosine potential and extreme starting angle. While we leave a full MCMC analysis to future work, we provide a simple comparison of these models to an approximate likelihood (using a $\chi^2$ comparison to the eBOSS DR14 Ly-$\alpha$ forest estimates of the linear MPS) to illustrate the potential constraining power of the data on the extreme axion model.

\subsection{Changes to axion background variables}

Before the onset of oscillations, when the axion field is slowly evolving over the negatively-curved potential, the axion perturbation equations of motion depend only on the derived background variables $w_\mathrm{ax}$ and $c_\mathrm{ad}^2$, defined in Eq. \ref{eqn:wax} and Eq. \ref{eqn:cad2}. In this section, we will discuss the effect of changing the axion starting angle and axion mass on the evolution of these background variables, as they help shed light on how the extreme starting angles impact the fluid equations of motion during these early times.

\begin{figure}
	\centering
    \includegraphics[trim={0.3cm 0.6cm 1.2cm 2.0cm},clip,width=0.5\textwidth]{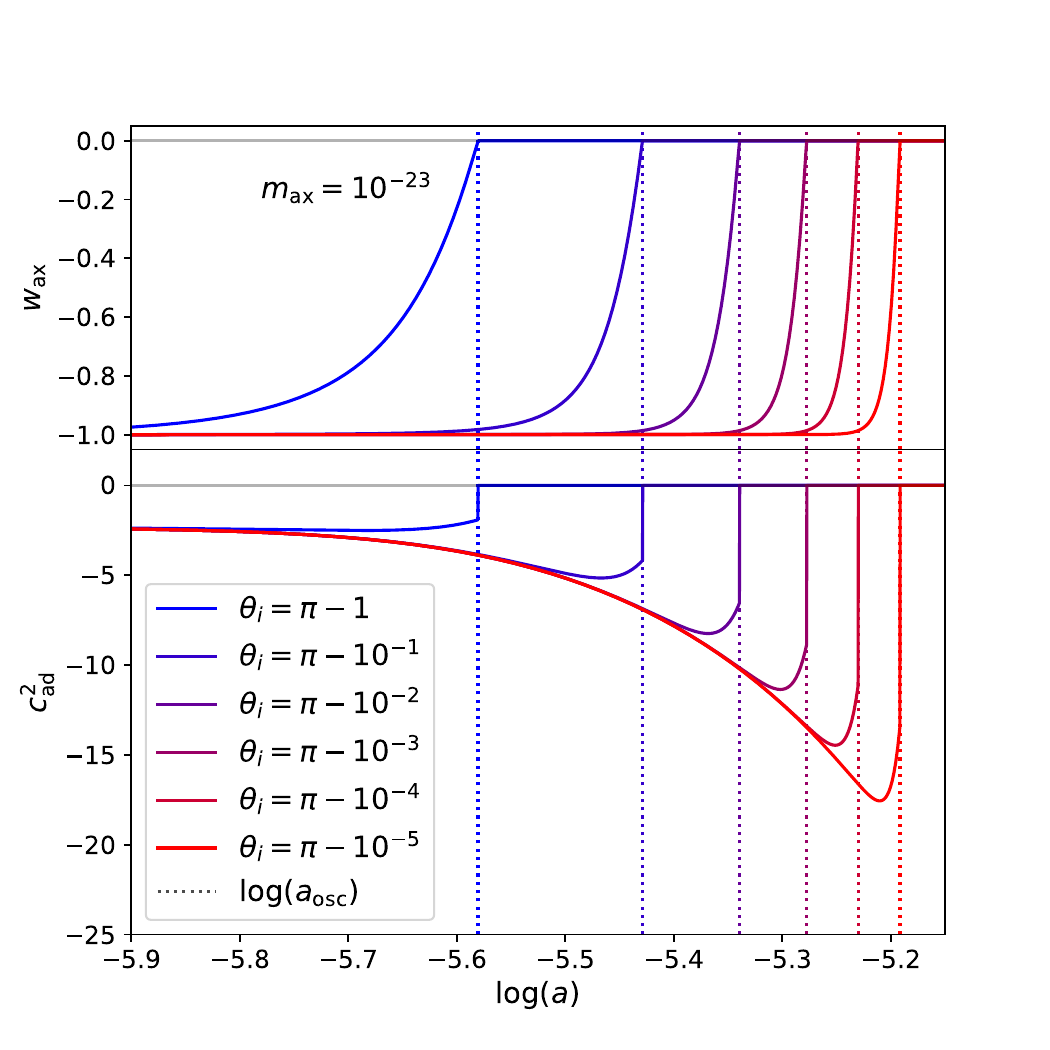}
   \caption{This plot shows the effect of varying the axion starting angle $\theta_i$ on the evolution of the axion background variables, $w_\mathrm{ax}$ and $c_\mathrm{ad}^2$, in the case of an axion mass of $10^{-24}$ eV. The dotted vertical lines represent the onset of axion oscillations, denoted as the first time when $w_\mathrm{ax} = 0$. We can see that for extreme starting angles, close to $\pi$ (in red), a number of features can be seen. The onset of oscillations is delayed, $w_\mathrm{ax}$ approaches zero more rapidly just before the onset of oscillations, and $c_\mathrm{ad}^2$ becomes much more negative before returning to zero.}  
    \label{fig:background_theta}
\end{figure}

\begin{figure}
	\centering
    \includegraphics[trim={0.5cm 1.0cm 1cm 2.2cm},clip,width=0.5\textwidth]{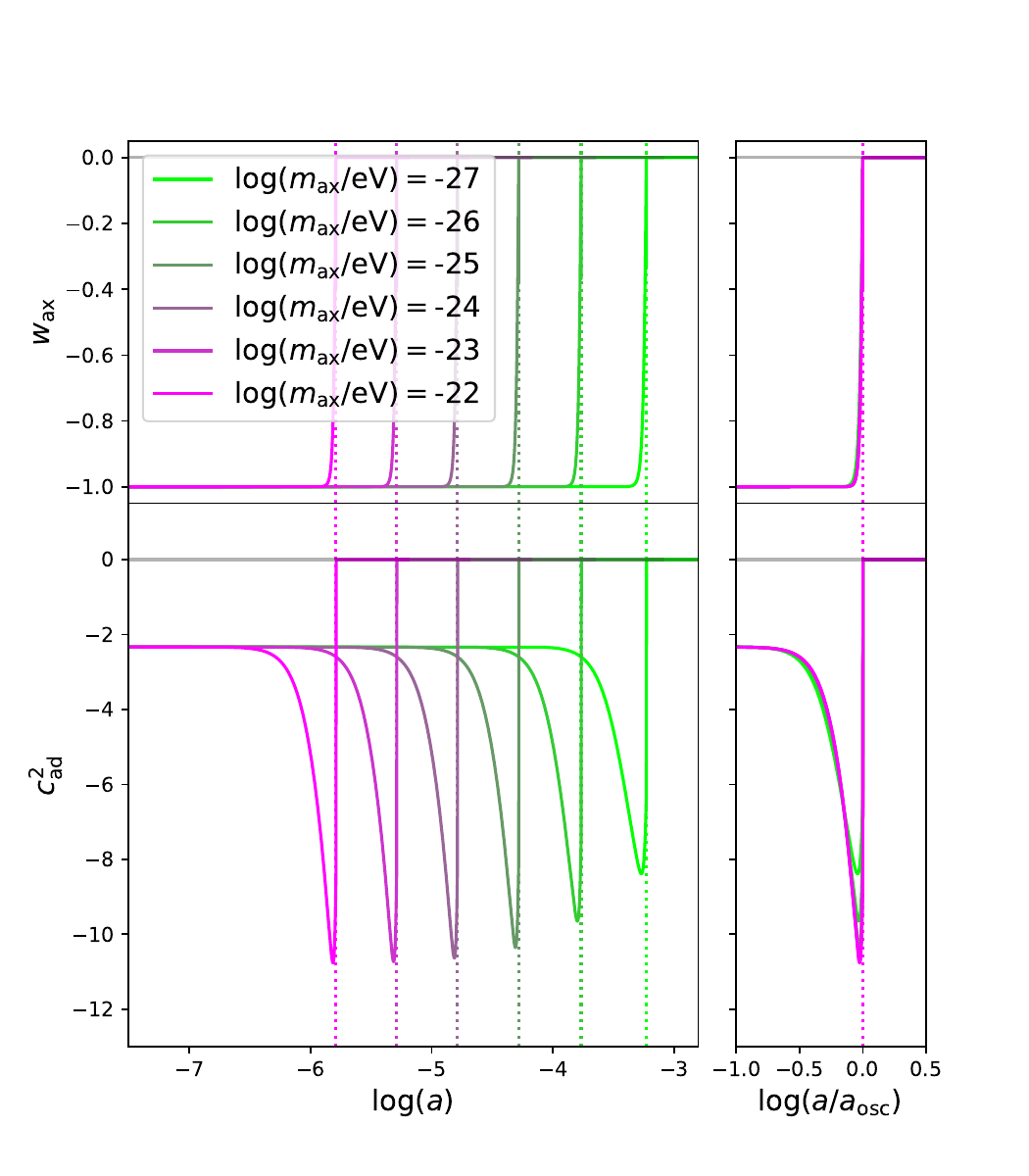}
   \caption{This plot shows the effect of varying the axion mass $m_\mathrm{ax}$ on the evolution of the axion background variables, $w_\mathrm{ax}$ and $c_\mathrm{ad}^2$, with a moderately extreme axion starting angle of $\theta_i = 3.14$. \rev{The right plots show the axion background variables normalized to the scale of the onset of oscillations ($a_\mathrm{osc}$).} We can see that the shape of the background variables changes only slightly with mass, and the largest change is a delayed onset of oscillations for low mass (in green).}  
    \label{fig:background_mass}
\end{figure}

There has not been a thorough treatment of extreme axions in the fluid formalism previously. The impacts of starting angle on the background evolution of $w_\mathrm{ax}$ and $c_\mathrm{ad}^2$ can be seen in Figure~\ref{fig:background_theta}. For low starting angles (blue lines near $\theta_i=1.0$ on the plot), the equation of state parameter $w_\mathrm{ax}$ starts at $-1$ at early times, as the fixed axion field behaves like a cosmological constant at these times. However, as the axion field starts to roll within the potential, the equation of state parameter rises from $-1$, crossing zero at the point when oscillations are defined to begin.
However, for extreme axion starting angles (in purple and red in Figure~\ref{fig:background_mass}), the onset of these oscillations in $w_\mathrm{ax}$ is delayed to later times, due to a flatter initial potential slope when the field starts near the cosine peak. Once the field does start to evolve, the evolution to $w_\mathrm{ax}$ rises to zero much faster, due to the fact that the Hubble friction has been allowed to become lower by the time the field starts to evolve. Therefore, as soon as the field enters the steeply sloped region of the potential, there is less holding it back from oscillating rapidly.

The adiabatic sound speed ($c_\mathrm{ad}^2$), shown in the lower subplot of Figure~\ref{fig:background_mass}, exhibits some of the same features as the equation of state parameter $w_\mathrm{ax}$, but with some notable differences. Like $w_\mathrm{ax}$, the evolution of $c_\mathrm{ad}^2$ starts at the same negative value for all starting field angles (in this case, starting at $-7/3$, as predicted by Ref. \onlinecite{Hu:2000ke}), evolving up to zero at the onset of oscillations, for the low field angles in blue. For extreme starting angles (purple/red lines with e.g. $\theta_i = \pi - 10^{-8} $) also have a delayed onset of evolution in the adiabatic sound speed, similar to those seen in $w_\mathrm{ax}$. However, one interesting new feature seen in $c_\mathrm{ad}^2$ for extreme starting angles is that the value of $c_\mathrm{ad}^2$ becomes extremely negative just before the onset of oscillations. This can be understood since $c_\mathrm{ad}^2 = \dot{P}_\mathrm{ax}/\dot{\rho}_\mathrm{ax}$ (Eq. \ref{eqn:cad2}), and when the field starts to slowly roll along the top of a cosine potential, the density $\rho_\mathrm{ax}$ is not decreasing as quickly as it would if the field were evolving down the side of a quadratic potential. This smaller denominator results in a larger absolute value for the adiabatic sound speed. 

These features of the smooth background variables can help explain the tachyonic growth of structure in the fluid formalism, as an extremely negative value of $c_\mathrm{ad}^2$ just before the onset of oscillations drives growth in certain terms in the fluid density perturbation equations of motion (Eq. \ref{eqn:pert_EoM}).

We can also examine how the axion mass changes the evolution of these key background fluid variables in the case of an extreme axion, as seen in Figure~\ref{fig:background_mass}. This plot also shows the equation of state parameter $w_\mathrm{ax}$ in the upper subplot, and the adiabatic sound speed $c_\mathrm{ad}^2$ in the lower subplot, for a fairly extreme starting angle of $\theta_i = 3.14$ and a range of axion masses. We can see that the shape of the evolution of these background variables is largely independent of mass, other than the onset of oscillations begins later (at larger values of the scale factor) for more massive axions.

\begin{figure}
	\centering
    \includegraphics[trim={0.2cm 0.7cm 1.2cm 1.0cm},clip,width=0.5\textwidth]{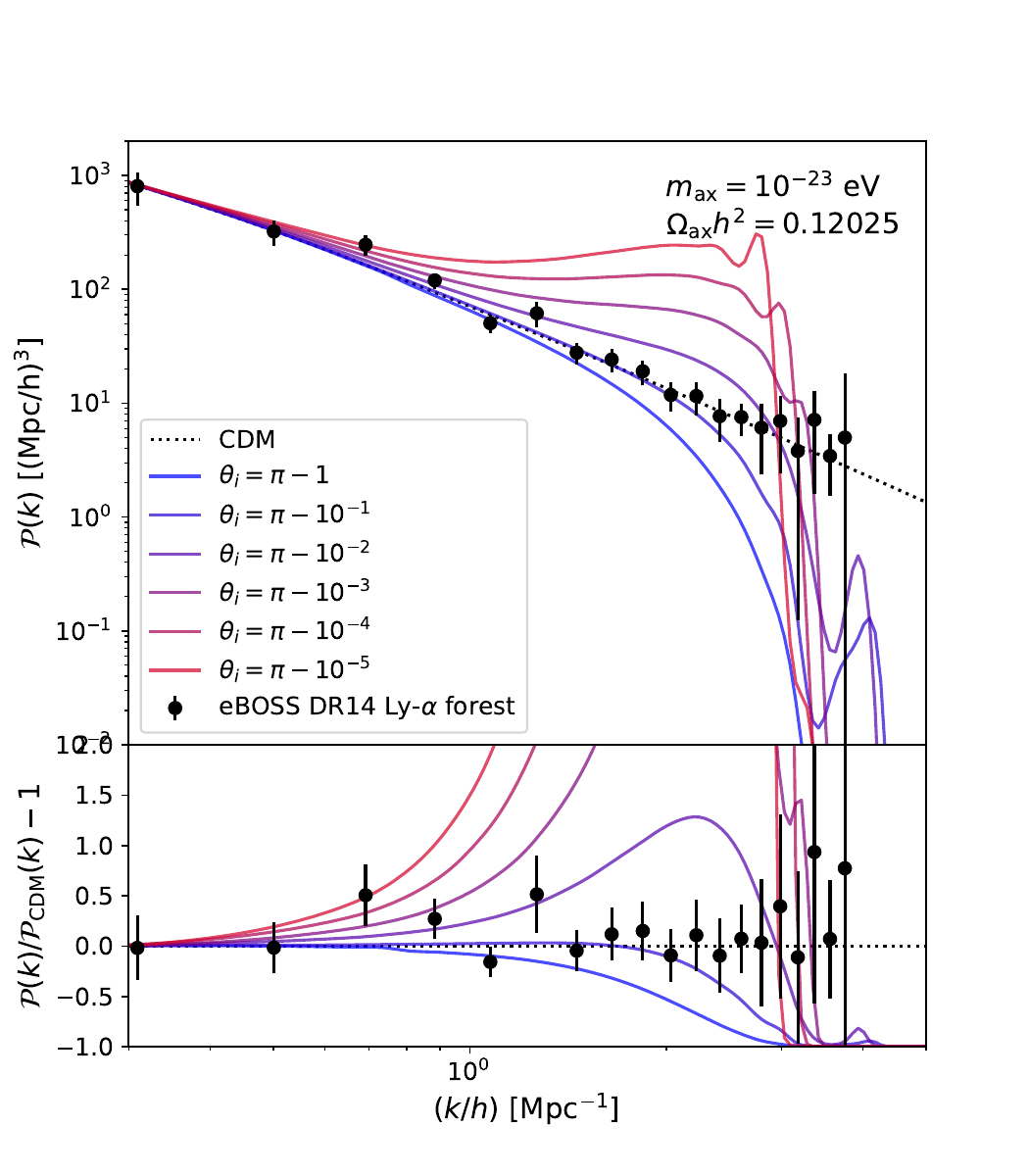}
   \caption{The effect of varying the axion starting angle ($\theta_i$) on the axion matter power spectrum.  The black data points with error bars are the published MPS data from the eBOSS DR14 Ly-$\alpha$ forest results. These models shown were computed with an axion mass of $m_\mathrm{ax} = 10^{-23}$ eV, and an axion density of $\Omega_\mathrm{ax}h^2 = 0.12025$, or constituting 100\% of the dark matter.}
    \label{fig:mps_theta}
\end{figure}

\begin{figure}
	\centering
    \includegraphics[trim={0.2cm 1.0cm 1.2cm 1.0cm},clip,width=0.5\textwidth]{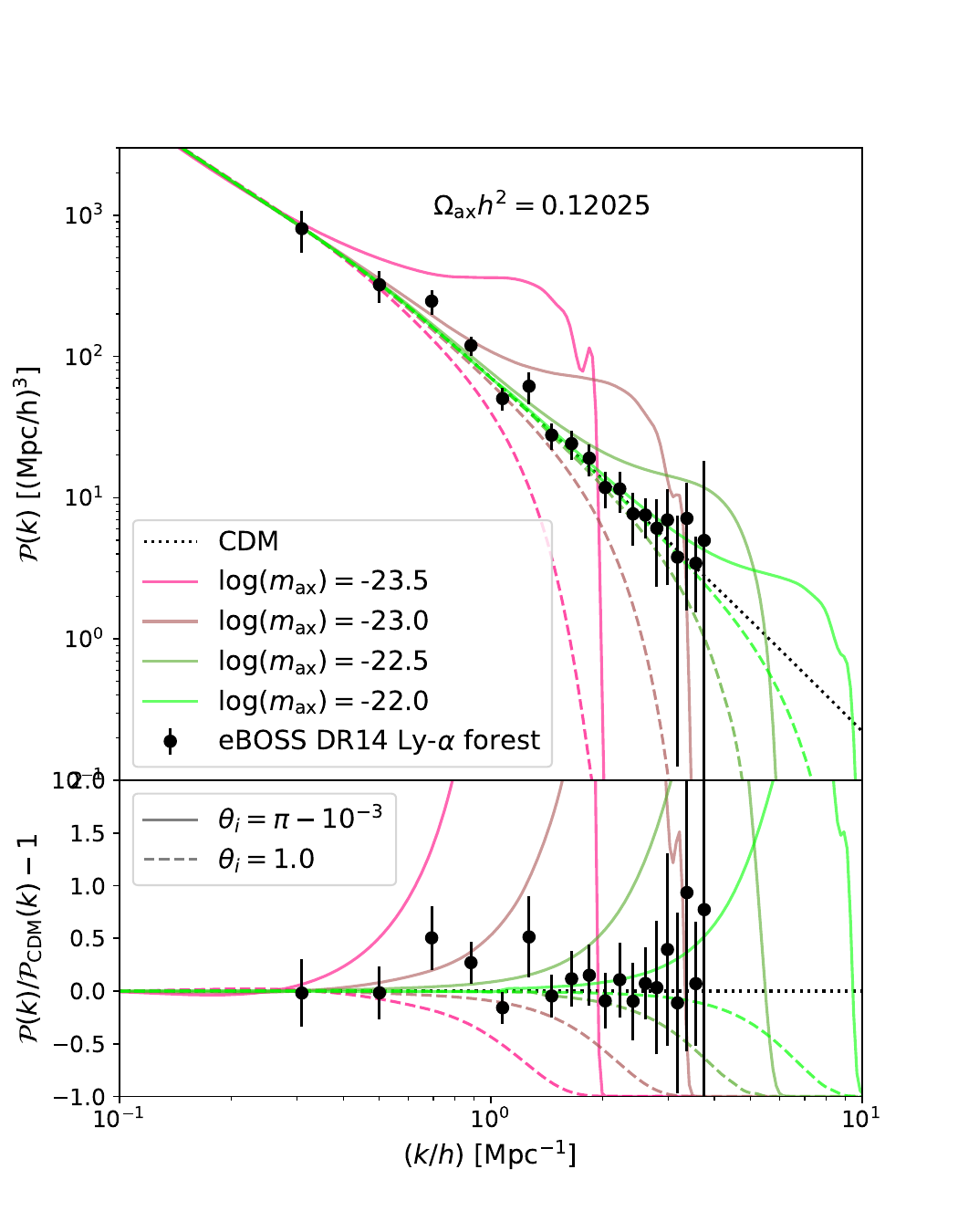}
   \caption{The effect of varying the axion mass ($m_\mathrm{ax}$) on the axion matter power spectrum, for both a low and extreme axion starting angle ($\theta_i = 1.0$ in dashed and $\pi - 10^{-8}$ in solid).  As in Figure~\ref{fig:mps_theta}, the black data points with error bars are the linear MPS estimates from the eBOSS DR14 Ly-$\alpha$ forest data. These results were computed with an axion density of $\Omega_\mathrm{ax}h^2 = 0.12025$, or constituting 100\% of the dark matter.}  
    \label{fig:mps_mass}
\end{figure}

\begin{figure}
	\centering
    \includegraphics[trim={0.2cm 0.3cm 1.2cm 1.0cm},clip,width=0.5\textwidth]{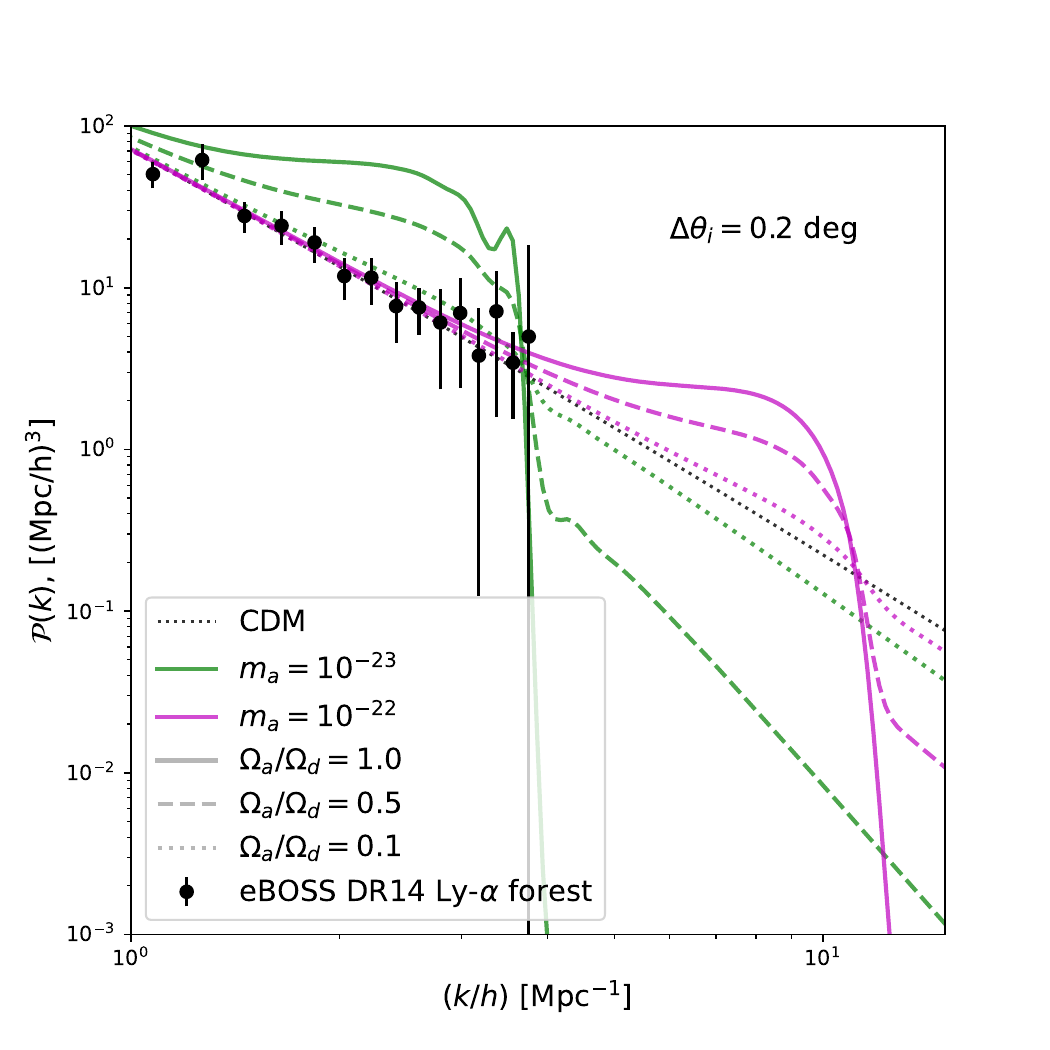}
   \caption{The effect of varying the axion fraction $(\Omega_\mathrm{ax} / \Omega_d)$ for two axion masses and a fixed extreme starting angle of $\Delta \theta_i = 0.2$ deg. As in Figures~\ref{fig:mps_theta} and \ref{fig:mps_mass}, the black data points with error bars are the linear MPS estimates from the eBOSS DR14 Ly-$\alpha$ forest data.}  
    \label{fig:mps_frac}
\end{figure}

\subsection{Matter Power Spectrum Signatures}

These changes to the background fluid variables also impact the MPS, which is what gives us the cosmological observables that can be seen in Ly-$\alpha$ forest. In this section, we describe the impact of axion mass, starting angle, and axion DM fraction on the MPS, and compare the results to the linear MPS estimated using the eBOSS DR14 Ly-$\alpha$ forest data and the $\Lambda$CDM model, as described in Section \ref{sec:lya}. Note that we are not doing a full hydrodynamical simulation of the Ly-$\alpha$ flux power spectrum, but are instead using the $z=0$ linear matter power spectrum estimated using the Ly-$\alpha$ forest data. This estimation has a number of limitations. The linearization of the Ly-$\alpha$ power spectrum, and the evolution to $z=0$, both assume pure CDM physics. In addition, these estimates marginalise over a number of astrophysical parameters describing the nonlinear fluid dynamics, which may have non-trivial degeneracies with both cosmological and axion parameters, which would need to be investigated more thoroughly in a robust comparison to Ly-$\alpha$ forest data. Therefore, this comparison should not be considered quantitatively robust, but instead as a qualitative demonstration of how and where extreme axions can alleviate previous Ly-$\alpha$ forest constraints on vanilla axion models. 

Figure~\ref{fig:mps_theta} shows the matter power spectrum for a variety of starting axion field angles, all for a fixed axion mass ($m_\mathrm{ax} = 10^{-23}$ eV) and fixed dark matter density ($\Omega_\mathrm{ax} h^2 = 0.12025$). We can see here that for a low starting angle well within the quadratic regime ($\theta_i = \pi - 1$, in blue) there is a reduction in power at small scales~\cite{Khlopov:1985jw}, which drives well-known limits on the fuzzy dark matter particle mass \cite{PhysRevD.91.103512,Marsh:2015xka,Irsic:2017yje,Rogers:2020ltq,Dentler:2021zij}. However, when we go to extreme starting angles (with $\theta_i \rightarrow  \pi$, in red) we can see an enhancement in power around the cutoff scale, eventually even surpassing the CDM results in black \citep[consistent with the the results of][]{Cedeno:2017sou,Cembranos:2018ulm,Zhang:2017dpp,Zhang:2017flu,Leong:2018opi,Arvanitaki:2019rax}. The eBOSS DR14 Ly-$\alpha$ forest estimates of the linear MPS are plotted in black with error bars, for visual comparison to the axion power spectra. We can see by eye that while both extremely low and extremely high starting angles appear to be strongly ruled out by the data, there is a range of starting angles around $\theta_i \approx \pi - 10^{-2}$ that agree with the data more, suggesting that for a certain axion mass and energy density, the starting angle may be able to be constrained from both sides.

Figure~\ref{fig:mps_mass} shows how the matter power spectrum depends on axion mass, for both low and high axion starting angle, again overlaid with the eBOSS DR14 Ly-$\alpha$ forest data. The axion mass changes the cutoff scale in the matter power spectrum for the low-angle vanilla axions, with lower mass axions exhibiting a reduction in power at larger scales (lower $k$ values), in agreement with Ref. \onlinecite{PhysRevD.91.103512}. The axion mass also changes the scale at which enhancement in the matter power spectrum occurs for the extreme axions. In a similar manner as the vanilla axion cutoff, the extreme axion enhancement occurs at larger scales (smaller $k$ values) for lower axion mass. The two effects appear to be synchronized, with a similar shift in $k$ for both the vanilla cutoff and the extreme enhancement. By comparing the eBOSS DR14 Ly-$\alpha$ forest data to the models we can see that measurements at smaller scales allow us to constrain both the vanilla and extreme axion models at higher masses.

Figure~\ref{fig:mps_frac} shows how the MPS depends on the axion fraction, for two masses and a fixed extreme starting angle. As expected, lower axion fractions result in the MPS converges to the CDM solution, suggesting that any extreme axion model can be unconstrained at a low enough axion DM fraction.

\begin{figure*}
	\centering
    \includegraphics[width=\textwidth]{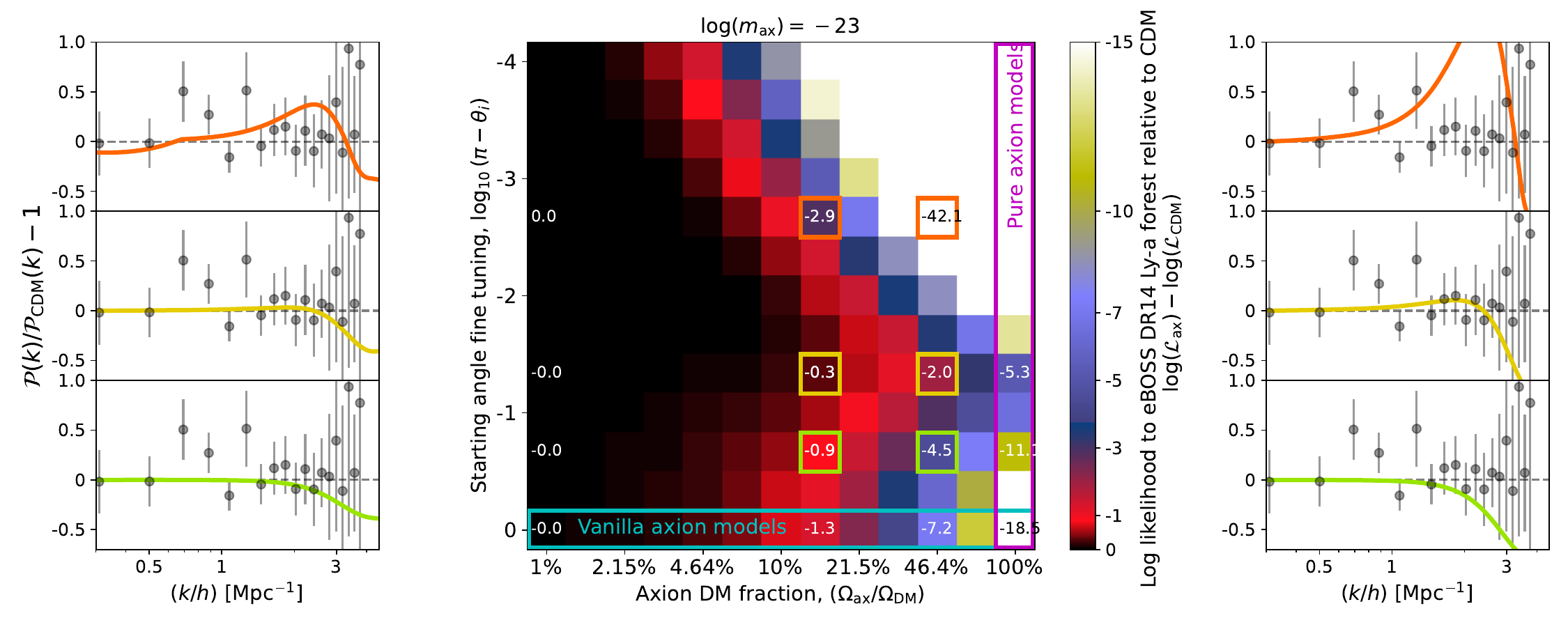}
   \caption{Simple fits of our linear MPS predictions to estimates from the eBOSS DR14 Ly-$\alpha$ forest, for a range of axion fractions and axion starting angles, and with a mass of $\log_{10}(m_\mathrm{ax}/\mathrm{eV}) = -23$. The middle plot shows a grid of the relative log likelihood values of this fit (using the chi-squared likelihood given in equation \ref{eq:likelihood}), relative to the fit with pure CDM (value close to zero on the colorbar means a fit that is almost as good as CDM, while a more negative value means a worse fit). The horizontal axis shows different axion fractions, on a logarithmic scale, so the models on the far right (within the magenta box) are pure axion models with no CDM. The vertical axis shows different degrees of fine-tuning for the axion starting angle. The row along the bottom (in the cyan box) have very low starting angles, so these results approximate those of a vanilla quadratic axion, while the row along the top have starting angles separated from $\pi$ by just $10^{-4}$ radians. The six plots along the left and right show the fractional differences of the matter power spectra for six examples on the grid, highlighted by the orange, yellow, and green boxes, to allow a visual comparison to the eBOSS DR14 Ly-$\alpha$ forest estimates of the linear MPS.}  
    \label{fig:chisgrid_23}
\end{figure*}

\subsection{Comparison to Ly-$\alpha$ forest estimates of the MPS}

 These MPS results can be compared to the eBOSS DR14 Ly-$\alpha$ forest estimates of the linear matter power spectrum and these estimates can be used to compute a simple $\chi^2$ to gauge the goodness of fit. 
The results of these likelihood comparisons are shown for $m_\mathrm{ax} = 10^{-23}$ eV in Figure~\ref{fig:chisgrid_23}. In this figure, we plotted the difference in log likelihoods between our extreme axion fits and a CDM fit, so that this difference should approach zero for extremely low axion fractions as we approach a pure CDM universe. In this case, the `likelihood' being computed was a simple $\chi^2$ metric to the eBOSS DR14 estimates of the linear MPS for model $m$, where the log likelihood is given by the equation,
\begin{equation}
    \label{eq:likelihood}
    \log(\mathcal{L}_m) \approx \chi^2_m = \sum_i \frac{(\mathcal{P}_i - \mathcal{P}_m(k_i))^2}{\sigma_{\mathcal{P}i}^2},
\end{equation}
where the eBOSS DR14 estimates of the MPS are $\mathcal{P}_i$ at wavenumber $k_i$ with uncertainty $\sigma_{\mathcal{P}i}$. This chi-squared likelihood can be computed for both an axion model and $\Lambda$CDM, and the difference of the logarithm of these results is plotted in Figure~\ref{fig:chisgrid_23}. A low value means the axion model is almost as good a fit as CDM, while higher values mean the axion model has an increasingly worse fit to the eBOSS data. As expected, the best fits are for low axion fractions, indicating that CDM is still the best fit to the eBOSS data. However, if we compare models with different starting angles and a fixed axion fraction, we can see that for fractions above $\sim 10\%$, varying the starting angle can result in a significantly better fit to the Ly-$\alpha$ forest estimates of the MPS.

In order to compare multiple different models, each with a different maximum likelihood and number of free parameters, it is useful to use the Akaike Information Criterion (AIC), given by
\begin{equation}
    \text{AIC} = 2K - 2\ln(\hat{L}),
\end{equation}
where $K$ is the number of free parameters in the model, and $\hat{L}$ is the maximum likelihood of the fit to a certain set of data \cite{Akaike:1974}. The best model (able to achieve the best fit with the minimum number of parameters) is the model with the lowest AIC. Therefore, when comparing two different models, the improvement is considered worth the added complexity if the difference in the natural logarithm of the maximum likelihood ($\Delta \log(\hat{L})$) is greater than the number of added parameters in the model. Here, we are using a $\chi^2$ comparison to the linearized matter power spectrum estimate (eq. \ref{eq:likelihood}) instead of a full likelihood computation of the Ly-$\alpha$ forest data, and we are also not maximizing the likelihood over all possible cosmological parameters. Despite these caveats, the AIC can still help to assess the improvement in our fit to the MPS estimates, and compare that improvement to the number of added parameters necessary to achieve that improvement.

When comparing extreme axion models to the standard axion implementation, one extra parameter is needed (either the axion starting angle $\theta_i$, or the potential scale parameter $f_\mathrm{ax}$, depending on how you formulate the problem). This added complexity to the model is justified if it yields $\Delta \log(\hat{L}) > 1$. Although we are not computing the maximum $\hat{L}$ marginalized over all cosmological parameters, we can still see the difference in $\log(L)$ in Figure~\ref{fig:chisgrid_23}, and we can see that considering extreme axions can improve the relative $\log(L)$ by several orders of unity, suggesting that the improvement in fit will be worth the extra parameter to the axion model.
The extreme starting angles can also significantly improve the fits for a range of other axion fractions. Although pure CDM is still a better fit to the eBOSS MPS estimates than axions with this mass of $m_\mathrm{ax} = 10^{-23}$ eV, we can see that moderately extreme axions (with a starting angle of $\theta_i \approx \pi - 10^{-1}$) offer a significantly better fit than vanilla axions for a range of axion fractions, while for very extreme starting angles ($\theta_i \gtrsim \pi - 10^{-3}$) the results once again are in tension with the eBOSS estimates.

\begin{figure*}
	\centering
    \includegraphics[width=\textwidth]{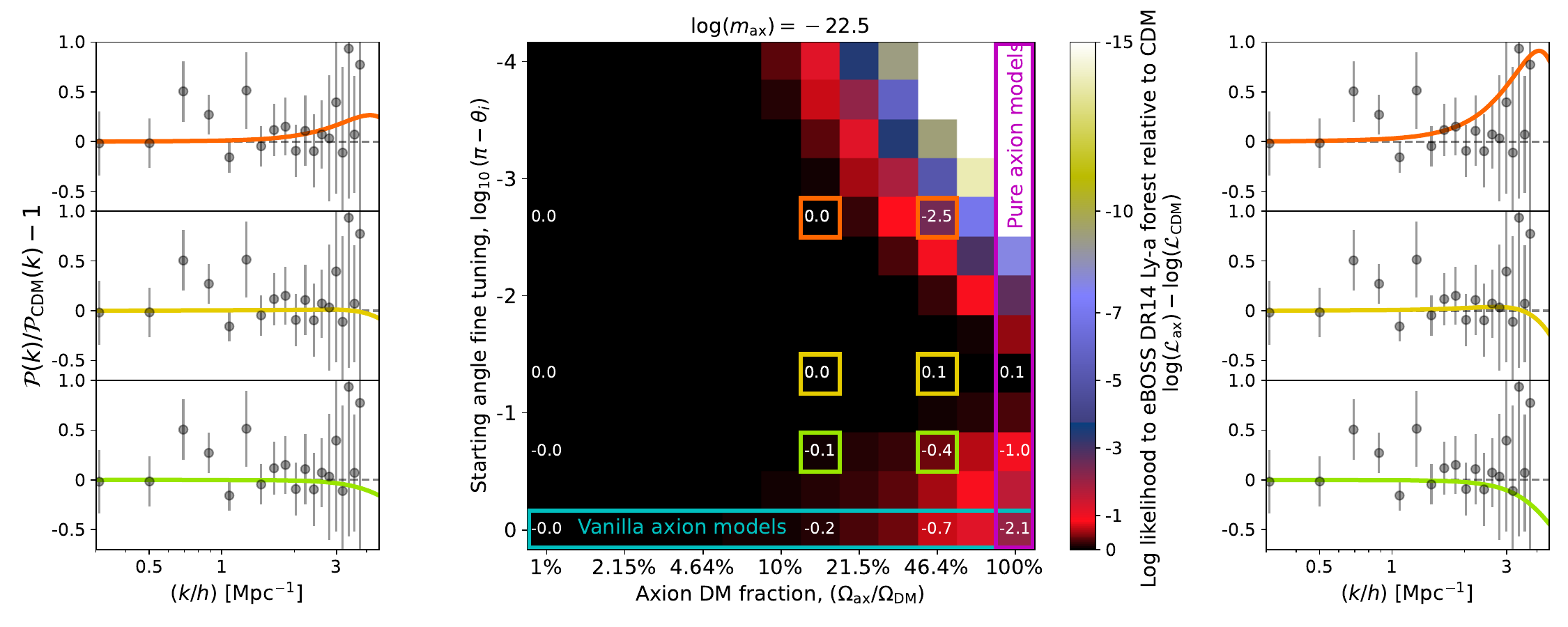}
   \caption{As Fig.~\ref{fig:chisgrid_23}, except for $\log_{10}(m_\mathrm{ax}/\mathrm{eV}) = -22.5$. The middle plot shows a grid of the relative log likelihood values of this fit, relative to the fit with pure CDM  The horizontal axis shows different axion fractions, and the vertical axis shows different degrees of fine-tuning for the axion starting angle. The six plots along the left and right show the fractional differences of the matter power spectra for six examples on the grid, highlighted by the orange, yellow, and green boxes, to allow a visual comparison to the eBOSS DR14 Ly-$\alpha$ forest estimates of the linear MPS. \rev{The best fits to Ly-$\alpha$ forest data can be achieved with a starting angle roughly between $3 \lesssim \theta_i \lesssim 3.13$, but fits can be achieved with lower axion fraction for a wider range of angles than that.}}  
    \label{fig:chisgrid_225}
\end{figure*}

\begin{figure}
	\centering
    \includegraphics[trim = {1.3cm, 1.3cm, 1.0cm, 0.7cm}, clip, width=\linewidth]{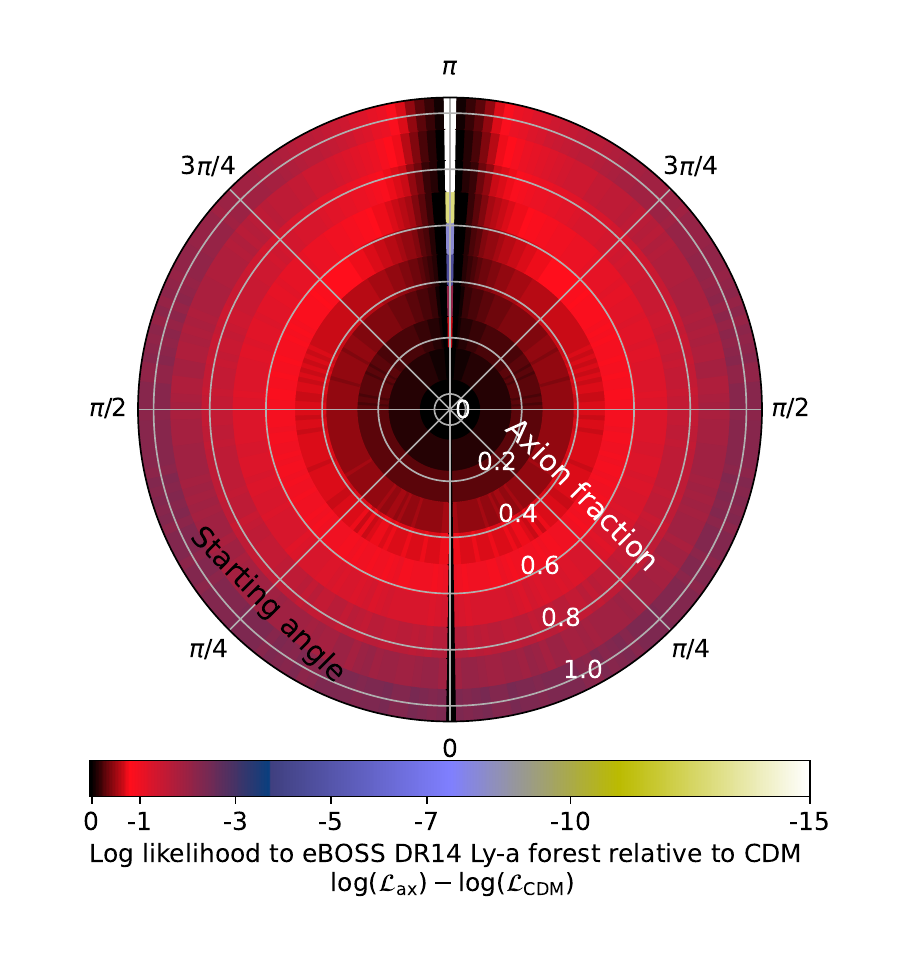}
   \caption{\rev{As Fig.~\ref{fig:chisgrid_225}, with $\log_{10}(m_\mathrm{ax}/\mathrm{eV}) = -22.5$, but plotted radially for a full range of starting angles. The color represents the relative log likelihood values of this fit, relative to the fit with pure CDM  The radial axis shows different axion fractions, and angle represents the axion starting angle. The black region represents scenarios that can fit the data as well as CDM, which include a non-trivial fraction of starting angles at high axion fraction.}}  
    \label{fig:chisgrid_circ}
\end{figure}

This mass of $m_\mathrm{ax} = 10^{-23}$ eV has a dramatic difference between vanilla and extreme axions, due to the large amount of overlap between the scales measured by the eBOSS Ly-$\alpha$ forest, and the scales affected by the extreme axion enhancements. However, even the maximum likelihood for pure axions, with a starting angle of $\theta_i = \pi - 10^{-1}$, still gives a difference to CDM of $\Delta \log{\hat{L}} \sim 5$, indicating that is still a poor fit to the Ly-$\alpha$ forest data, and a significantly worse fit than CDM. We can get better agreement, however, if we go to slightly higher masses. 

Figure~\ref{fig:chisgrid_225} shows the same log likelihood comparison for a range of axion fractions and starting angles, but this time for a mass of $m_\mathrm{ax} = 10^{-22.5}$ eV. Once again, we can see that pure CDM still gives the best fit, but for a fixed axion fraction, axions with a starting angle between $3 \lesssim \theta_i \lesssim 3.13$ have higher likelihoods than vanilla axions (along the lowest row). In particular, for pure axion models (along the far right), the maximum likelihood around $\theta_i = \pi - 10^{-1} \approx 3.01$ gives log likelihood values that are actually $0.1$ higher than CDM, whereas vanilla axions are $2.1$ less than CDM. This suggests that extreme axions with $m_\mathrm{ax} = 10^{-22.5}$ eV may actually be slightly preferred by the Ly-$\alpha$ data, while vanilla axions with the same mass would be ruled out. In addition, the improvement is likely significant enough to warrant the addition of one extra parameter, based on the AIC. This conclusion could be verified with a full hydrodynamical simulation and comparison to Ly-$\alpha$ forest data, and the exact mass where axions can go from forbidden to permitted might change slightly. However, we can be confident that extreme axions can alleviate tensions with Ly-$\alpha$ data for some mass range, drawing into question our previous upper bounds on the axion mass, and motivating future work into Ly-$\alpha$ constraints on extreme axions.

\rev{The range of possible starting angles that can relieve axion constraints is non-trivial. Fig.~\ref{fig:chisgrid_circ} shows the relative difference in log likelihood for $m_\mathrm{ax} = 10^{-21.5}$ eV relative to CDM for a full range of starting angles, where the polar angle is the axion starting field angle, and the radius is the axion fraction. The dark region of the plot (with fits that are almost as good as CDM) covers the low-fraction region in the centre of the plot, as well as a non-trivial portion of the outer region around starting angles close to $\pi$. From this plot, we can conclude that even with a uniform prior on starting angle, the existence of these extreme solutions should have a nontrivial impact on estimated constraints on axion mass and fraction.}

\begin{figure*}
	\centering
    \includegraphics[width=\textwidth]{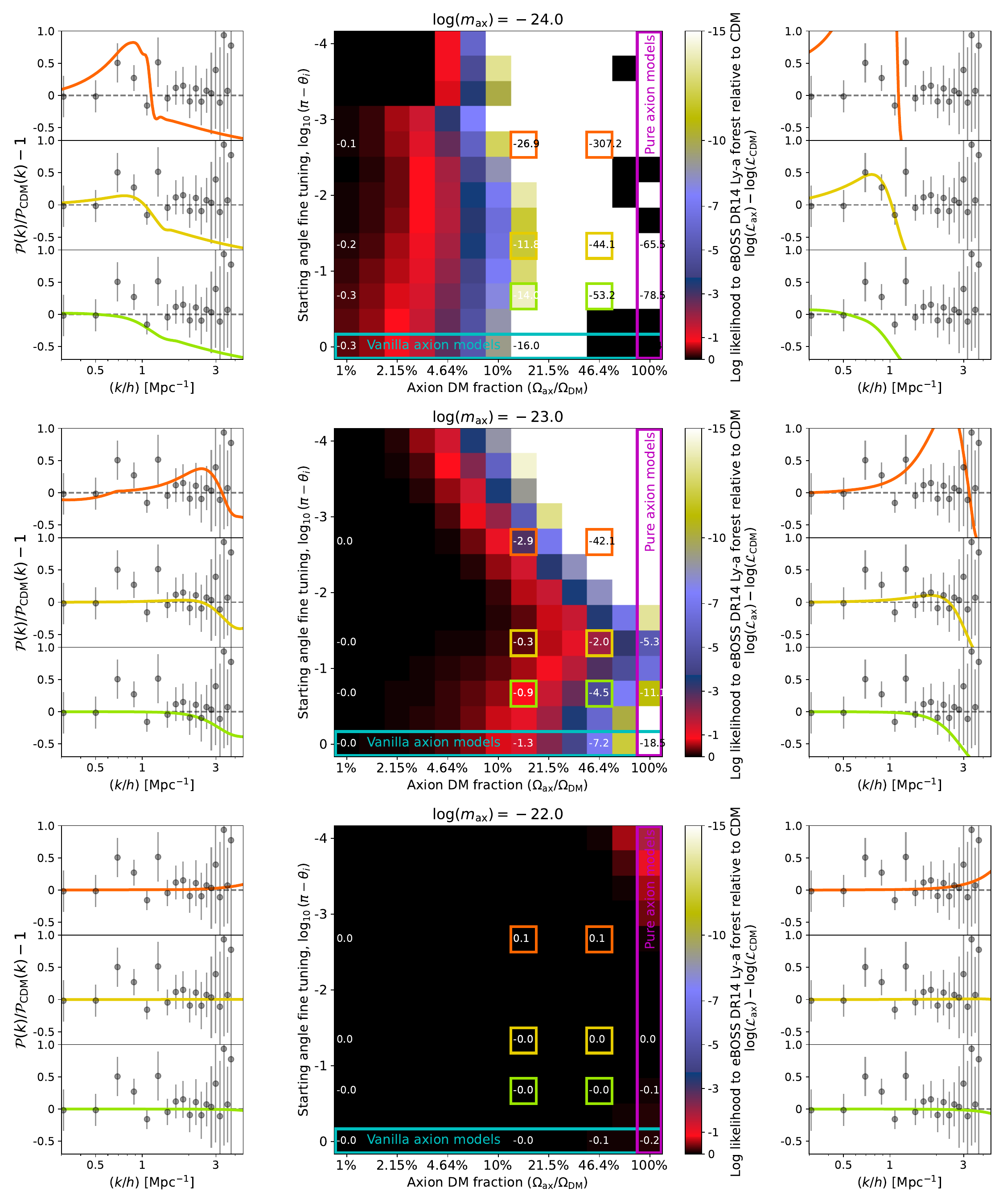}
   \caption{The fit of our extreme mixed axion model to the eBOSS Ly-$\alpha$ forest data for a larger range of masses, showing that axions with masses of $10^{-24}$ eV or $10^{-22}$ eV are not impacted by considering extreme axions as opposed to vanilla axions.}  
    \label{fig:chisgrid_3mass}
\end{figure*}

While Figures \ref{fig:chisgrid_225} and \ref{fig:chisgrid_23} show that a moderately extreme axion can help alleviate tensions with Ly-$\alpha$ forest data for axions with a mass of $10^{-22.5}$ eV, this is not necessarily true for all axion masses. Figure~\ref{fig:chisgrid_3mass} shows this same log likelihood grid for masses of $10^{-22}$ and $10^{-24}$ eV. We can see that for $m_\mathrm{ax} = 10^{-24}$ eV, the effects are at low enough $k$ that they remain ruled out regardless of initial field angle. In this case, the constraints become entirely dependent on the axion fraction. On the other hand, at $m_\mathrm{ax} = 10^{-22}$ eV, the effects are at high enough $k$ that they are completely unconstrained by the eBOSS Ly-$\alpha$ forest estimates of the MPS, regardless of starting field angle or axion density fraction. Evidently, extreme axion starting angles can only alleviate constraints on the axion density fraction with Ly-$\alpha$ forest data for a specific range of masses, around $m_\mathrm{ax} \sim 10^{-23}$ eV. It is important to note that the relevant mass range in question will depend heavily on the maximum $k$ being probed by the relevant survey. This mass range will change slightly if we use Ly-$\alpha$ estimates at different scales \citep[such as high-resolution surveys using the Keck telescopes or the Very Large Telescopes (VLT),][]{Lu:1996sn, Irsic:2017sop}, but the effect will likely still be limited to a certain mass range.

\section{Discussion and Future Work}
\label{sec:discussion}

The methods presented in this work allow for rapid computations of cosmological observables in extreme axion models, and their rapid comparison to real data, and opens up a range of interesting applications and areas of further study. In general, moderately extreme starting angles seem to alleviate tensions with existing measurements of the MPS for a certain range of axion mass $m_\mathrm{ax}$, axion density $\Omega_\mathrm{ax} h^2$, and initial axion field angle $\theta_i$, but establishing the exact limits of these alleviated tensions is an interesting question that would require an extensive MCMC analysis. \rev{We can see from Fig.~\ref{fig:chisgrid_circ} that even with uniform priors on the starting angle (i.e. no fine-tuning mechanism), the existence of these extreme solutions capable of alleviating the constraints would make up a non-trivial fraction of model parameter space.} Such an analysis would involve repeated computations of our extreme axion model (made possible by the rapid run times of our modified \texttt{axionCAMB}), and the repeated comparison to cosmological likelihoods for galaxy clustering and the CMB. Such an analysis would tell us exactly how much weaker our constraints on axion mass and density could be if we are allowed to vary the starting field angle. It would also tell us about any degeneracies between these three axion parameters and any other cosmological parameters being varied.

Although comparison to LSS likelihoods from galaxy surveys, and CMB likelihoods for the lensing, temperature, and polarization power spectra are the most straightforward, the tightest current constraints on axions come from measurements of the Ly-$\alpha$ forest, as these are able to probe the MPS at much smaller scales than either galaxy surveys or the CMB \cite{Rogers:2020ltq}. However, comparing MPS predictions for extreme axions to data from the Ly-$\alpha$ forest is more difficult, as it requires hydrodynamical simulations of the small-scale nonlinear structure, which in principle could depend on the nonlinear behaviour of the extreme axion model. In this paper, we used the estimates of the linear $z=0$ MPS from the Ly-$\alpha$ forest data, which assumed CDM for the small-scale structure evolution, but this method is only valid in the low-axion-density regime, where CDM makes up most of the dark matter.\footnote{\rev{Ref.~\cite{Rogers:2023upm} finds that pure CDM, with the hydrodynamical model used in the eBOSS analysis, is insufficient to explain both the \textit{Planck} CMB and eBOSS Ly-$\alpha$ forest data, further motivating a more robust mixed-axion hydrodynamical treatment of the Ly-$\alpha$ forest.}} Some work has been done modeling the nonlinear Ly-$\alpha$ forest for extreme axions\cite{Leong:2018opi}, but this simulation is computationally expensive. Ideally, the best approach would be to train an emulator to produce extreme axion predictions of the Ly-$\alpha$ data, similar to what was done in Ref. \onlinecite{Rogers:2020ltq}. When combined with our modified \texttt{axionCAMB}, this could allow for rapid computation and direct comparison to Ly-$\alpha$ forest data, which would give the most informative constraints on the small-scale behaviour of these extreme axion models. \rev{This analysis would also require a more robust model of the extreme axion sound speeds in order to model the effects at higher redshifts, while the current model has only been validated for $z=0$.} In addition, direct comparison to Ly-$\alpha$ observables would allow us to use higher resolution spectroscopic surveys, such as those done with Keck or VLT \cite{Lu:1996sn, Irsic:2017sop}.

Accurate simultaneous constraints on the axion mass, density fraction, and starting angle, would quantitatively address an important question that, so far, has only been approached qualitatively: namely, the required degree of fine-tuning for these extreme axion models to work. Fig.~\ref{fig:chisgrid_225} shows that a good agreement with data can be reached with axion starting angles that are close to the peak, separated by less than 10\%. \rev{Fig.~\ref{fig:chisgrid_circ} shows that the range of starting angles that can fit the Ly-$\alpha$ forest data with a large axion fraction is non-trivial, suggesting that extreme starting angles are an important factor to consider when computing axion constraints, even without a fine-tuning mechanism.}
This required degree of fine-tuning could also depend on other cosmological parameters. 
With our modified \texttt{axionCAMB}, we could create estimates of the necessary degree of fine-tuning for a range of axion and cosmological parameters, helping to inform the plausibility of these models that produce starting angles close to $\pi$.

Another area worth exploring is comparing these constraints to forecast sensitivities by future CMB experiments, such as the Simons Observatory, and CMB-S4 \cite{Hlozek:2016lzm, 2019BAAS...51g.147L, Dvorkin:2022bsc, CMB-S4:2022ght}. Although Planck is already cosmic-variance limited for temperature at low-$\ell$, there may be substantial improvements to be made with an experiment with better polarization and/or high-$\ell$ data \cite{Aghanim:2015xee}. CMB lensing also offers the ability to probe the DM MPS at a range of scales\cite{Rogers:2023ezo}. We could also experiment with simultaneous constraints from CMB and MPS sources. Direct probes of the MPS can also be used to constrain the extreme axion model, including the Dark Energy Survey~\citep[which we used to constrain the vanilla axion model in][]{Dentler:2021zij}, \emph{Euclid}~\cite{Amendola:2016saw}, JWST \cite{Parashari:2023cui}, and the Vera Rubin Observatory \cite{Mao:2022fyx}. Weak gravitational lensing or Galactic dynamics could also be used to search for DM substructure on sub-galactic scales \cite{Mondino:2020rkn, Winch:2020cju, Mondino:2023pnc}.

\rev{It is important to note that these ultralight axion models can also be probed by late-time astrophysical effects, including the formation of dwarf galaxies \citep{Dalal:2022rmp}, or dark matter rotation curves \citep{Bar:2019bqz, Bar:2021kti}. These late-time astrophysical probes would not be impacted by the early-time imprints of extreme starting angle, and thus these constraints appear to be independent of axion starting angle. However, these constraints depend on astrophysical modeling in dense baryonic environments (for example, modeling the structure of the soliton core), which makes it difficult to probe to low axion fraction. Cosmological constraints sensitive to the linear MPS (such as the CMB, Ly-$\alpha$ forest, or the UV luminosity function) remain superior in probing low axion fractions \cite[eg.][]{PhysRevD.91.103512, Rogers:2023ezo, Winch:2024mrt}, and thus a rigorous understanding of extreme starting angles is necessary to compute accurate constraints using these experiments. While these observables probe different axion masses (ranging from $10^{-27}$ eV up to $10^{-19}$ eV), we anticipate that the alleviation of mass and fraction constraints due to extreme axion starting angle will be qualitatively similar across that range. The comparison to eBOSS Ly-$\alpha$ forest constraints is intended as an illustrative proof of concept, and is not intended to be a comprehensive reevaluation of cosmological axion constraints.}

Lastly, there is scope to constrain potentials beyond just those with the standard cosine shape. Models have been proposed with axions with quartic, hyperbolic cosine, or monodromic potentials \cite{Cembranos:2018ulm, Urena-Lopez:2019xri, Jaeckel:2016qjp, LinaresCedeno:2021sws}. In addition, axion-like scalar fields with a variety of potentials have been proposed as an early dark energy component potentially capable of relieving the Hubble tension \cite{Kamionkowski:2022pkx, Poulin:2023lkg}. Axion perturbations in all of these potentials could conceivably be modeled using our modified \texttt{axionCAMB}, since the potential function is implemented generically. The only requirement would be that the potential being tested must simplify to a quadratic at small $\phi$ values, in order for the particle DM approximation to be valid at late times.

\section{Conclusions}

Extreme axions represent an interesting class of dark matter models, theoretically motivated by string theory, and with distinct signatures on cosmological observables. Previously, their one major drawback was the high computational cost of modeling the rapid field oscillations. In this work, we have introduced a new extension to the existing \texttt{axionCAMB} software, allowing it to compute MPS and CMB observables for extreme axion models in $\sim 7$ seconds, where previous models have taken multiple days. These observables can be computed for a range of values for the axion mass, axion DM density fraction, and extreme axion starting angle, as well as a range of ordinary cosmological parameters. We achieved this rapid modeling of the extreme axions by using a modified version of \texttt{axionCAMB}'s fluid approximation, re-configuring the initial conditions to allow for finely tuned starting angles, modifying the effective fluid sound speed to reflect the tachyonic growth during the oscillatory phase, and implementing an efficient lookup table of the axion background fluid variables to allow for rapid computation. 

We also compared the results of our extreme axion model to estimates of the linear matter power spectrum from the eBOSS DR14 Ly-$\alpha$ forest data. While there are limitations to this approach, as the estimation of the linear $z=0$ MPS from Ly-$\alpha$ forest data assumes CDM physics, and integrates over a number of astrophysical parameters, we can still use this comparison to give us estimates of the effect of these extreme axion models \rev{on cosmological axion constraints more generally}. We find that when considering the eBOSS DR14 Ly-$\alpha$ forest data, for a range of axion masses around  $m_\mathrm{ax} \approx 10^{-22.5} \mathrm{\ eV}$, constraints on the axion fraction can be significantly weakened by considering extreme axions with a starting angle between $\pi - 10^{-1} \lesssim \theta_i \lesssim \pi - 10^{-2}$. This motivates future work running robust MCMC comparisons of this extreme axion model to Ly-$\alpha$ observables, as well as CMB and other cosmological axion measurements. With the help of this new, efficient fluid model of extreme axions, we can compute more nuanced constraints on axion mass and fraction, as well as shed new light on the possible high-energy origins of these ALPs through estimates of the axion decay constant $f_\mathrm{ax}$.


\section{Acknowledgements}
We are grateful to Wayne Hu, Tzihong Chiueh, Asimina Arvanitaki, Alex Lagu\"e, and Rayne Liu for their helpful discussions and advice.

The Dunlap Institute is funded through an endowment established by the David Dunlap family and the University of Toronto. The authors at the University of Toronto acknowledge that the land on which the University of Toronto is built is the traditional territory of the Haudenosaunee, and most recently, the territory of the Mississaugas of the New Credit First Nation. They are grateful to have the opportunity to work in the community, on this territory.

Computations were performed on the Niagara supercomputer at the SciNet HPC Consortium. SciNet is funded by Innovation, Science and Economic Development Canada; the Digital Research Alliance of Canada; the Ontario Research Fund: Research Excellence; and the University of Toronto.

HW would like to acknowledge the support of the Natural Sciences and Engineering Research Council of Canada (NSERC) Canadian Graduate Scholarships - Master's  and Canadian Graduate Scholarships - Doctoral programs. DJEM is supported by an Ernest Rutherford Fellowship from the Science and Technologies Facilities Council (Grant No. ST/T004037/1) and by a Leverhulme Trust Research Project (RPG-2022-145). RH acknowledges support from the NSERC.  RH additionally acknowledges support from CIFAR, and the Azrieli and Alfred P. Sloan Foundations.

\bibliography{main_bibliography}

\end{document}